\RequirePackage{ifpdf}
\documentclass[12pt]{article}
\usepackage{epsfig,graphicx,bm,amsmath,amsfonts}
\usepackage{epstopdf}
\usepackage{amssymb,xcolor,float}
\usepackage{amsthm}
\usepackage{hyperref}
\usepackage{cite}
\input epsf.sty
\usepackage{cite}
\usepackage{comment}
\usepackage{epstopdf}
\usepackage{float}
\usepackage{caption}
\usepackage{subcaption}
\usepackage{enumitem}
\usepackage{bbm}
\usepackage{bm}
\usepackage{longtable}
\usepackage{parskip}
\usepackage{enumitem}
\usepackage{graphicx,epsfig}
\usepackage{ae,aecompl}
\usepackage{dcolumn}
\usepackage{latexsym}
\usepackage{psfrag}
\usepackage{ytableau}
\usepackage{tabularx}
\usepackage{float}
\usepackage{wrapfig}

\usepackage{xcolor}
\usepackage[english]{babel}
\usepackage[T1]{fontenc}
\usepackage[utf8]{inputenc}
\usepackage{authblk}
\usepackage{mathtools}
\usepackage{slashed}
\usepackage{mattens,amssymb}
\usepackage{amsmath,amssymb}
\usepackage{amsfonts}
\usepackage{graphicx,color}
\usepackage{cite}
\usepackage{hyperref}
\hypersetup{
	colorlinks=true,
	linkcolor=blue,
	filecolor=magenta,      
	urlcolor=cyan,
}
\usepackage[capitalize]{cleveref}
\numberwithin{equation}{section} 

\definecolor{refcol}{rgb}{0.9,0.1,0.1}
\hypersetup{colorlinks=true,linkcolor=blue,citecolor=refcol,urlcolor=cyan,linktocpage}

\usepackage{wrapfig}

\newcommand{\Tr}{\text{Tr}}

\newcommand{\nn}{\nonumber}

\newcommand{\ben}{\begin{eqnarray}\displaystyle}
\newcommand{\een}{\end{eqnarray}}

\newcommand{\be}{\begin{equation}}
\newcommand{\ee}{\end{equation}}

\newcommand{\bc}{\begin{center}}
\newcommand{\ec}{\end{center}}

\newcommand{\eesp}{\end{split}}
\newcommand{\bsp}{\begin{split}}

\newcommand{\Rmnum}[1]{\expandafter\@slowromancap\romannumeral #1@}
\renewcommand{\b}{\beta}		%%% Redefinition
\renewcommand{\t}{\tau}		%%% Redefinition
\newcommand{\cB}{\mathcal{B}}
\newcommand{\cC}{\mathcal{C}}

\newcommand{\cH}{\mathcal{H}}

\newcommand{\cP}{\mathcal{P}}

\newcommand{\cS}{\mathcal{S}}

\newcommand{\cZ}{\mathcal{Z}}

\def\tL{\text{L}}

\newcommand{\where}{\text{where}}

\newcommand{\tand}{\text{and}}

\newcommand{\bensp}{\begin{eqnarray}\begin{split}}
\newcommand{\eensp}{\end{eqnarray}\end{split}}

\newcommand{\bnm}{\begin{matrix}}
\newcommand{\enm}{\end{matrix}}
\def\XXint#1#2#3{{\setbox0=\hbox{$#1{#2#3}{\int}$ }
\vcenter{\hbox{$#2#3$ }}\kern-.6\wd0}}

\newcommand{\ket}[1]{|#1\big>}
\newcommand{\bra}[1]{\big<#1|}

\newcommand{\braket}[2]{\big<#1|#2\big>}

\newcommand{\gb}{\beta}

\newcommand{\gz}{\zeta}
\newcommand{\gs}{\sigma}
\def\cZ{\mathcal{Z}}
\def\cB{\mathcal{B}}
\def\cP{\mathcal{P}}
\def\cS{\mathcal{S}}

\def\t{\text}

\begin{document}

\begin{center}
{\Large \bf{Bosonization, BTZ Black Hole Microstates, and Logarithmic Correction to Entropy}}
\end{center}
\baselineskip=18pt

\bigskip

\centerline{Suvankar Dutta$^{a}$, Shruti Menon$^{a}$ and Aayush Srivastav$^{b}$}

\bigskip

\centerline{\large \it $^{a}$Indian Institute of Science Education and Research Bhopal}
\centerline{\large \it Bhopal bypass, Bhopal 462066, India}
\centerline{\large \it $^{b}$Indian Institute of Science Education and Research Kolkata }
\centerline{\large \it Mohanpur, Nadia 741246, India}

\bigskip

\centerline{E-mail: suvankar@iiserb.ac.in, shruti98.menon@gmail.com, aayush1309physics@gmail.com}

\vskip .6cm
\medskip

\vspace*{4.0ex}

\begin{abstract}

We study three-dimensional gravity with negative cosmological constant under non-standard boundary conditions where chemical potentials are determined dynamically. Using a boundary Hamiltonian inspired by collective field theory (ColFT), the boundary dynamics reduce to those of a one-dimensional fluid on a circle, with configurations corresponding to bulk geometries such as BTZ black holes. Quantizing the system via bosonization of relativistic fermions, we obtain a microscopic description of black hole states in terms of Young diagrams, whose degeneracies match the Bekenstein--Hawking entropy.

We compute the Euclidean canonical partition function and free energy for both the ColFT Hamiltonian and a relativistic free-fermion Hamiltonian. In the ColFT case, the partition function resembles that of chiral \( U(N) \) Yang--Mills theory on a torus, with \( N \sim 1/(\beta G) \). This offers a novel way to compute quantum corrections to the partition function. The leading entropy term receives contributions from all genera, while the subleading logarithmic correction is one-loop exact, arising solely from the genus-one sector with coefficient \( -\frac{1}{2} \). This coefficient remains unchanged in the relativistic fermion case, suggesting the universality of the one-loop correction across different boundary Hamiltonians.

\end{abstract}

\vfill\eject

\tableofcontents

\section{Introduction}

Three-dimensional spacetime -- regardless of the presence of a cosmological constant -- provides a remarkably powerful and tractable setting for studying the classical and quantum properties of gravity. This is largely due to the fact that, in three dimensions, classical gravity possesses no local degrees of freedom in the bulk; all dynamics are determined entirely by the boundary conditions imposed at infinity. In particular, when the cosmological constant is negative, Brown and Henneaux showed that an appropriate choice of boundary conditions leads to an asymptotic symmetry group generated by two copies of the Virasoro algebra~\cite{Brown:1986nw}. Although these Brown–Henneaux boundary conditions are widely considered standard, many alternative sets of boundary conditions have been studied rigorously in the literature~\cite{Henneaux:2013dra, Bunster:2014mua, Perez:2016vqo, Perez:2012cf, Afshar:2016wfy, Grumiller:2016pqb, Grumiller:2016kcp, Grumiller:2019tyl, Afshar:2016kjj, Afshar:2013vka, Troessaert:2013fma, Avery:2013dja, Ammon:2017vwt, Ozer:2019nkv, Gonzalez:2018jgp, Campoleoni:2010zq}. Exploring this broader landscape of boundary conditions can offer significant insights into the fundamental nature of gravity.

In this context, $\mathrm{AdS}_3$ gravity supports a novel class of non-standard boundary conditions wherein the chemical potentials—i.e., the time components of the gauge fields—are not held fixed at the boundary. Instead, they are allowed to depend explicitly on the fields, in particular the angular components of the gauge fields. This dependence is governed by a chosen boundary Hamiltonian. Following the framework developed in~\cite{Dutta:2023uxe}, we adopt a boundary Hamiltonian identified with the collective field theory (ColFT) Hamiltonian originally proposed by Jevicki and Sakita~\cite{sakita, jevicki}, and refer to the corresponding boundary condition as the \emph{ColFT boundary condition}. Defined on a cylinder, the ColFT Hamiltonian arises from unitary matrix quantum mechanics with an arbitrary potential $W(\Tr U)$. Importantly, these boundary conditions are compatible with BTZ black hole solutions in the bulk.

The equations of motion for the collective field and its conjugate momentum closely resemble the continuity and Euler equations for a one-dimensional fluid on a circle with negative pressure. In this analogy, the collective field is interpreted as the fluid density. Consequently, the boundary dynamics of the gravitational theory can be understood in terms of fluid dynamics on the boundary of $\mathrm{AdS}_3$. Different fluid configurations correspond to distinct bulk geometries: a BTZ black hole (stationary and rotationally symmetric) maps to a fluid with constant density and velocity; an extremal BTZ black hole corresponds to a fluid whose velocity equals its density; and a static fluid represents a non-rotating black hole.

Quantization of this system proceeds by promoting the classical Poisson brackets to quantum commutators, which reveals that the asymptotic symmetry algebra becomes a $U(1)$ Kac–Moody algebra. We construct representations of this algebra using quantum bosonization of relativistic free fermions, establishing a connection between the fermionic degrees of freedom and the dynamical fields of $\mathrm{AdS}_3$ gravity. This construction enables us to define the Hilbert space of the fermionic system and interpret specific bulk configurations—viewed as bosonic excitations—as particular particle--hole pair excitations above the \(\mathbf{n}\)-particle ground state \(\ket{\mathbf{n}}\) in the fermionic Hilbert space. In this framework, the microstates of a BTZ black hole with fixed mass and angular momentum are identified with such particle--hole pair excitations.

We find that these microstates can be labeled by Young diagrams associated with irreducible representations of $U(N)$, with the number of boxes in the diagram determined by the parameters of the black hole. By computing the degeneracy of these states, we show that the logarithm of the degeneracy exactly reproduces the classical Bekenstein–Hawking entropy.

We further develop the Euclidean canonical partition function for the BTZ black hole and compute the free energy for two distinct choices of boundary Hamiltonians. We begin with the ColFT Hamiltonian, for which the partition function in each sector closely resembles that of the chiral sectors of a two-dimensional \(U(N)\) Yang--Mills theory on a torus. The rank \(N\) is identified with \(\mathbf{n}\) of the \(\mathbf{n}\)-particle ground state \(\ket{\mathbf{n}}\) of the Hilbert space, and it scales as \(\mathbf{n} \sim 1/(\beta G)\), where \(G\) is Newton's constant and \(\beta\) denotes the size of the Euclidean time circle. By evaluating this partition function explicitly, we compute both the leading and subleading contributions to the entropy.

The Yang--Mills partition function admits a genus (or topological) expansion. In the classical limit (\(\beta G \ll 1\)), the leading contribution to the entropy arises from the full topological expansion, receiving contributions from all genera. In contrast, the subleading logarithmic correction originates solely from the genus-one sector, rendering it one-loop exact. In particular, we find that the coefficient of the logarithmic correction is \(-\frac{1}{2}\).

The leading term in the free energy consists of two parts. The first arises from the genus-one contribution and correctly reproduces the free energy of the BTZ black hole relative to a suitable background geometry. The second part, which receives contributions from all other genera, is of the same order as the genus-one term. We interpret this additional contribution as arising from other saddle points in the path integral, although we have not yet identified these saddles explicitly.

We repeat the analysis using an alternative choice of boundary Hamiltonian corresponding to relativistic fermions, which also admits a BTZ black hole in the bulk. This computation is technically simpler, and at leading order, the free energy exactly reproduces that of the BTZ black hole. In contrast to the ColFT case, contributions from additional saddle points are absent. Remarkably, the logarithmic correction remains unchanged, with a coefficient of \(-\frac{1}{2}\), highlighting the universality of this one-loop contribution across different boundary conditions.

The structure of the paper is organized as follows. In Section~\ref{sec:Ads3gravity}, we provide a brief overview of \(AdS_3\) gravity and introduce the non-standard boundary conditions relevant to our analysis. Section~\ref{sec:colHamBTZ} presents the boundary Hamiltonian associated with the collective field theory (ColFT) description. In Section~\ref{sec:BTZ}, we derive the BTZ black hole geometry that emerges from imposing ColFT boundary conditions and discuss how a similar geometry arises when the boundary Hamiltonian corresponds to relativistic fermions. Section~\ref{sec:quantisation} is devoted to the quantization of the system: we begin with the bosonization of relativistic fermions, construct the Hilbert space, and identify the quantum states corresponding to a given black hole geometry. In Section~\ref{sec:canonical}, we formulate the canonical partition function for the BTZ black hole and compute both the leading and subleading contributions to the free energy in the classical limit. Finally, Section~\ref{sec:conclusion} summarizes our main results and outlines open questions and directions for future research.

\paragraph{Note added:} This work constitutes a significant extension of our previous results reported in \cite{Dutta:2025ypr}, where we constructed the microstates associated with the BTZ black hole. In the present study, we advance this analysis by computing the full canonical partition function, thereby offering a deeper and more comprehensive understanding of the collective field theory (ColFT) boundary conditions for different boundary Hamiltonians.

\section{Gravity in $AdS_3$ and Chern-Simons theory}
\label{sec:Ads3gravity}

This section presents a compact summary of how three-dimensional Anti-de Sitter (AdS\(_3\)) gravity is formulated in terms of Chern–Simons theory. Gravity in the bulk of AdS\(_3\) spacetime can be described by a Chern–Simons gauge theory with gauge group \(SO(2,2)\)~\cite{Campoleoni:2010zq, Campoleoni:2011hg}. Notably, the Lie algebra \(so(2,2)\) is isomorphic to \(sl(2,\mathbb{R}) \oplus sl(2,\mathbb{R})\), allowing the gauge fields to be decomposed into two chiral components, denoted \(A^\pm\). Each of these matrix-valued gauge fields can be expressed in terms of the vielbein \(e^a_\mu\) and the spin connection \(\omega^{ab}_\mu\) as
\begin{align}\label{eq:gaugemetricrel}
	A^\pm = \left(\omega^a \pm \frac{e^a}{l}\right) T_a,
\end{align}
where \(l\) is the radius of the AdS\(_3\) spacetime, and \(T_a\) are the generators of \(SL(2,\mathbb{R})\). The dualized spin connection \(\omega_a\) is related to \(\omega^{bc}\) via
\begin{equation}
	\omega_a \equiv \frac{1}{2} \epsilon_{abc} \omega^{bc}.
\end{equation}

This decomposition enables a reformulation of the Einstein–Hilbert action in terms of the Chern–Simons action. Explicitly, the action is given by
\begin{align}\label{eq:CSacn}
	I = I_{CS}(A^+) - I_{CS}(A^-),
\end{align}
where the Chern–Simons action for each chiral component is
\begin{align}\label{eq:CSacn2}
	I_{CS}(A^\pm) = \frac{\mathrm{k}}{4\pi} \int \Tr\left[ A^\pm \wedge dA^\pm + \frac{2}{3} (A^\pm)^3 \right] + \cB_\infty(A^\pm).
\end{align}
Here, \(\mathrm{k}\) denotes the Chern–Simons level, which is related to the Newton constant \(G\) and the AdS radius \(l\) by
\begin{equation}\label{eq:kGrelation}
	\mathrm{k} = \frac{l}{4G}.
\end{equation}

In equation~\eqref{eq:CSacn2}, \(\cB_\infty(A^\pm)\) represents a boundary term included to ensure that the variation of the action \(\delta I_{CS}(A^\pm)\) vanishes. The trace is taken over the fundamental representation of the Lie algebra \(\mathfrak{sl}(2,\mathbb{R})\), whose generators are denoted \(\tL_\pm\) and \(\tL_0\), satisfying
\begin{align}
	\Tr( \tL_0 \tL_0) = \frac{1}{2}, \quad \Tr( \tL_1 \tL_{-1}) = -1,
\end{align}
with all other traces vanishing.

The spacetime metric can be reconstructed from the gauge fields using the relation
\begin{align}\label{eq:metric}
	g_{\mu\nu} = \frac{l^2}{2} \Tr((A^+ - A^-)_\mu (A^+ - A^-)_\nu).
\end{align}

In three-dimensional gravity, the theory is locally trivial and all physical dynamics emerge from the behavior at the boundary, which in turn is sensitive to the choice of boundary conditions.

\subsection{Boundary Conditions}

We introduce coordinates \((r, t, \theta)\) for the Lorentzian AdS\(_3\) manifold, where \(\theta\) is compact, \(r\) is the radial coordinate approaching the asymptotic boundary as \(r \to \infty\), and \(t\) denotes time. The gauge fields are parametrized as
\begin{equation}\label{eq:Apm-apm}
	A^\pm = b_\pm^{-1} \left( d + a^\pm \right) b_\pm,
\end{equation}
where \(b^\pm = b^\pm(r)\) are group elements that depend only on the radial coordinate, and the auxiliary connections \(a^\pm = a^\pm(t,\theta)\) depend only on the boundary coordinates \(t\) and \(\theta\).

To satisfy the bulk equations of motion (equivalent to the Einstein equations), the form of \(b^\pm\) need not be specified. We choose the temporal and angular components of \(a^\pm\) as~\cite{Grumiller:2016kcp}
\begin{equation} \label{eq:apmform}
	a^\pm(t,\theta) = \left( \xi_\pm(t,\theta) dt \pm p_\pm(t,\theta) d\theta \right) \tL_0,
\end{equation}
where \(p_\pm\) are dynamical variables, and \(\xi_\pm\) are interpreted as the chemical potentials.

The equations of motion derived from the Chern–Simons action are simply the flatness conditions
\begin{eqnarray}\label{eq:Maxeq}
	dA^\pm + A^{\pm 2} =0.
\end{eqnarray}
Substituting equations~\eqref{eq:Apm-apm} and \eqref{eq:apmform} into \eqref{eq:Maxeq}, we obtain the following relation between \(p_\pm\) and \(\xi_\pm\):
\begin{equation}\label{eq:maxeq}
	\Dot{p}_\pm(t,\theta) = \pm \xi_\pm'(t,\theta),
\end{equation}
where \(\dot{}\) and \( '\) denote derivatives with respect to \(t\) and \(\theta\), respectively.

To determine the boundary term \(\cB_\infty^\pm\), we require that the variation of the total action vanish:
\begin{eqnarray}\label{eq:Bterms}
	\delta \cB_\infty^\pm = - \frac{\mathrm{k}}{2\pi} \int dt \, d\theta \, \Tr( A_t^\pm \delta A^\pm_{\theta} ) = \mp \frac{\mathrm{k}}{4\pi} \int dt \, d\theta \, \xi_\pm \delta p_\pm.
\end{eqnarray}

In special cases where \(\xi_\pm\) are constants, the variation operator \(\delta\) may be pulled outside the integral, allowing direct computation of \(\cB_\infty^\pm\). However, in a more generic setup, one allows the chemical potentials to be field-dependent and defines \(\xi_\pm\) functionally in terms of some quantity \(H^\pm\) via
\begin{eqnarray}\label{eq:xidef}
	\xi_\pm = -\frac{4\pi}{\mathrm{k}} \frac{\delta H^\pm}{\delta p_\pm},
\end{eqnarray}
where \(H^\pm\) is expressed as a functional of \(p_\pm\) and its derivatives:
\begin{equation}\label{eq:boundHam}
	H^\pm = \int d\theta \, \cH^\pm(p_\pm, p_\pm', \cdots).
\end{equation}
Accordingly, the boundary term becomes
\begin{eqnarray}\label{eq:boundterm}
	\cB_\infty^\pm = \pm \int dt \, d\theta \, \cH^\pm = \pm \int dt \, H^\pm.
\end{eqnarray}

The dynamics of the boundary fields \(p_\pm\) are governed by the choice of \(\cH^\pm\). From the flatness condition, we obtain
\begin{eqnarray}\label{eq:boundarydynamics}
	\Dot{p}_\pm(t, \theta) = \mp \frac{4\pi}{\mathrm{k}} \frac{\partial }{\partial \theta} \left(  \frac{\delta H^\pm}{\delta p_\pm} \right).
\end{eqnarray}
We refer to \(H^\pm\) as the boundary Hamiltonians, and different choices correspond to different boundary dynamics.

These equations can also be derived from the Poisson structure:
\begin{equation}
	\dot p_\pm(t,\theta) = \{p_{\pm}(t,\theta),H^\pm\}_{PB},
\end{equation}
where the Poisson bracket is given by
\begin{equation}\label{eq:poisson}
	\{p_{\pm}(t,\theta), p_{\pm}(t,\theta')\}_{PB} = \mp \frac{4\pi}{\mathrm{k}} \frac{\partial}{\partial\theta} \delta(\theta-\theta').
\end{equation}

\section{Collective field theory and black holes in $AdS_3$ gravity}
\label{sec:colHamBTZ}

A novel set of boundary conditions for \(AdS_3\) gravity was introduced in~\cite{Dutta:2023uxe}, where the boundary dynamics of spin-two and higher-spin fields are governed by the interacting collective field theory (ColFT) Hamiltonian originally formulated by Jevicki and Sakita.

In this framework, the collective field and its conjugate momentum evolve analogously to a one-dimensional fluid with negative pressure. By adopting the ColFT Hamiltonian as the boundary Hamiltonian, one can interpret various bulk geometries in terms of fluid flows on a circle. Our primary interest lies in time-independent, spherically symmetric black hole solutions. It is found that black holes characterized by fixed mass and angular momentum correspond to a circular one-dimensional fluid with constant density and uniform velocity.

The collective field theory Hamiltonian is given by
\begin{equation}
	H_{CFT} = \int d\theta \ \sigma(t,\theta) \left[ \frac{1}{2} \left( \frac{\partial \Pi(t,\theta)}{\partial \theta}\right)^2 + \frac{\pi^2}{6} \sigma^2(t,\theta) + W(\theta) \right],
\end{equation}
where \(\sigma(t,\theta)\) denotes the eigenvalue density of an \(N \times N\) unitary matrix \(U\) in the large-\(N\) (continuum) limit, and \(\Pi(t,\theta)\) is the conjugate momentum associated with \(\sigma(t,\theta)\). The equations of motion for the collective variables are
\begin{eqnarray}
	\begin{split}
		\partial_t \sigma(t,\theta) + \partial_\theta (\sigma(t,\theta) v(t,\theta)) & = 0, \\
		\partial_t v(t,\theta) + v(t,\theta)\partial_\theta  v(t,\theta)  & = - \partial_\theta \left(\frac{\pi^2}{2} \sigma^2(t,\theta) + W(\theta)\right),
	\end{split}
\end{eqnarray}
where the velocity field \(v(t,\theta)\) is defined as
\begin{equation}
	v(t,\theta) = \partial_\theta \Pi(t,\theta).
\end{equation}
These correspond to the continuity and Euler equations for a one-dimensional fluid with negative pressure. In this picture, the eigenvalue density \(\sigma(t,\theta)\) is identified with the fluid density. This system forms a coupled set of nonlinear partial differential equations. To simplify the analysis, one can introduce the variables \(p_\pm(t,\theta)\), defined by
\begin{equation}\label{eq:ppmsv}
	p_\pm(t,\theta) = v(t,\theta) \pm \pi \sigma(t,\theta).
\end{equation}
In terms of these variables, the equations of motion decouple and reduce to
\begin{equation}
	\partial_t p_\pm(t,\theta) + p_\pm(t,\theta) \partial_\theta p_\pm(t,\theta) + W'(\theta) = 0,
\end{equation}
and the Hamiltonian becomes
\begin{equation}\label{eq:ColFTHam}
	H_{CFT} = H_{CFT}^+ + H_{CFT}^-, \quad \where \quad H_{CFT}^\pm = \pm \frac{1}{2\pi} \int d\theta \left(\frac{p^3_\pm(t,\theta)}{6} + W(\theta) p_\pm(t,\theta) \right).
\end{equation}
This Hamiltonian corresponds to the phase space Hamiltonian of non-relativistic free fermions on the circle \(S^1\)~\cite{Dutta:2023uxe}.

\subsection{Asymptotic Symmetries and Conserved Charges}

Asymptotic symmetries are identified as gauge transformations that preserve the asymptotic form of the gauge fields. Specifically, the asymptotic structure of the gauge connection defined in~\eqref{eq:apmform} remains invariant under the transformation
\begin{equation}\label{eq:gaugetrana}
	\delta a^\pm = d\lambda^\pm + [a^\pm, \lambda^\pm],
\end{equation}
where the gauge parameter is taken to be \(\lambda^\pm = \eta^\pm(t,\theta)\tL_0\). Under this transformation, the dynamical variables \(p^\pm\) and the chemical potentials \(\xi^\pm\) transform as
\begin{equation}\label{eq:gaugetranp}
	\delta p_\pm = \eta'_\pm, \quad \text{and} \quad \delta \xi_\pm = \dot\eta_\pm.
\end{equation}
These transformations do not arise from constraints and therefore constitute genuine asymptotic symmetries, despite being field-dependent and non-rigid~\cite{Banados:1998gg}. As such, they are associated with conserved charges that depend on the choice of gauge parameters \(\eta^\pm\).

Following the procedure outlined in~\cite{REGGE1974286, Banados:1994tn}, one can compute these conserved quantities. As shown in~\cite{Avan:1991kq, Avan:1991ik, Dutta:2023uxe}, there exists an infinite tower of conserved charges defined by
\begin{equation}\label{eq:Qn}
	Q^\pm_n = \pm\frac{\mathrm{k}}{4\pi} \int d\theta \sum_{k=0}^n \frac{^nC_k}{2^k(2k+1)} p^{2k+1}_\pm(t,\theta)W^{n-k}(\theta),
\end{equation}
where \(n = 1\) corresponds (up to an overall factor) to the Hamiltonian in~\eqref{eq:ColFTHam}. As implied by the Poisson bracket structure~\eqref{eq:poisson}, the charges \(Q^\pm_n\) with \(n \geq 2\) commute with the Hamiltonian and are therefore conserved.

Moreover, it can be explicitly verified that these conserved charges form an abelian algebra on-shell:
\begin{equation}
	\{Q^\pm_m, Q^\pm_n\}_{PB} = 0 , \quad \forall\ m,n \geq 1.
\end{equation}
This confirms the integrable structure of \(AdS_3\) gravity under the specified boundary conditions.

\section{BTZ black hole}\label{sec:BTZ}

In this section, we show that the collective field theory (ColFT) boundary conditions permit BTZ black hole solutions in the bulk. While our explicit analysis focuses on two particular choices of boundary Hamiltonians, the method we employ is general and readily extends to other Hamiltonians. We summarize the key steps and main results below, with full derivations provided in the appendix.

\subsection{Collective field theory Hamiltonian}\label{sec:BTZColFT}

We examine the boundary dynamics (\ref{eq:boundarydynamics}) governed by the Hamiltonian proportional to the ColFT Hamiltonian (\ref{eq:ColFTHam})
\begin{equation}\label{eq:HamCFT}
	H^\pm_{(1)} = \frac{\mathrm{\mathcal{C}_\pm k}}{2} H_{CFT}^\pm = \pm \frac{\mathcal{C}_\pm \mathrm{k}}{4\pi} \int d\theta \left(\frac{p^3_\pm(t,\theta)}{6} + W(\theta) p_\pm(t,\theta) \right).
\end{equation}
Here, we introduce a proportionality constant \(\mathcal{C}_\pm\), which will later be fixed to express the BTZ black hole metric in its standard form. While this choice is not strictly necessary, omitting it would require a rescaling of the time coordinate to achieve the same standard representation of the metric.

The chemical potentials $\xi^\pm$ can be calculated using the definition given in (\ref{eq:xidef})
\begin{equation}\label{eq:xip}
    \xi^\pm = \mp \mathcal{C}_\pm \left( \frac{p^2_\pm(t,\theta)}{2} + W(\theta) \right).
\end{equation}
For a time-independent geometry, the equations of motion given in (\ref{eq:maxeq}) imply that the chemical potentials \(\xi^\pm\) are constant, and the momenta \(p_\pm\) are independent of time \(t\). In this case, the equations reduce to  
\begin{equation}
	p_\pm(\theta)\partial_\theta p_\pm(\theta) + W'(\theta) = 0.
\end{equation}
This equation can equivalently be rewritten as
\begin{eqnarray}
	\frac{\partial}{\partial\theta} \left( v(\theta)\sigma(\theta) \right) = 0, \quad \text{and} \quad \frac{\partial}{\partial\theta} \left( \frac{\sigma^2(\theta) + v^2(\theta)}{2} \right) + W'(\theta) = 0.
\end{eqnarray}
When \( W'(\theta) \neq 0 \), both \(\sigma(\theta)\) and \(v(\theta)\) acquire \(\theta\)-dependence, corresponding to spherically non-symmetric solutions—commonly referred to as black flower geometries.

In this work, we concentrate on spherically symmetric black hole solutions. For such cases, \(W'(\theta) = 0\), and both \(\sigma(\theta)\) and \(v(\theta)\) become constant, greatly simplifying the analysis. The values of \(\sigma\) and \(v\) are then determined by specifying the black hole charges.

To ensure a smooth Euclidean geometry (i.e., no conical singularity in the Euclidean spacetime), one imposes that the holonomy of \(A^\pm\) around the Euclidean time circle is trivial up to conjugation:
\begin{equation}
    \text{Hol}_\tau(A^\pm) \sim \exp\left( \beta_\pm \xi^\pm \right) \sim \mathbb{I} 
    \quad \Rightarrow \quad 
    \beta_\pm \xi^\pm = \mp 2\pi.
\end{equation}
This requirement fixes the time components of the gauge fields \(\xi^\pm\) in terms of two parameters \(\beta_\pm\), periodicity of the Euclidean time in the ``\(+\)" and ``\(-\)" sectors. These relations in turn determine \(\sigma\) and \(v\) through equation~(\ref{eq:xip})
\begin{equation}\label{eq:vsigmabetapm}
    \sigma = \frac{\left(\beta _-+\beta _+\right) l}{\beta _- \beta
   _+}, \quad v =\frac{\pi l \left(\beta _--\beta _+\right)}{\beta _-
   \beta _+}.
\end{equation}
\iffalse
which encode how the gauge fields in each sector “wraps” around the Euclidean time circle. \(\beta_\pm\) do not correspond to two distinct time coordinates, but rather to how the ``\(+\)" and ``\(-\)" sectors of the gauge theory perceive the Euclidean thermal cycle. 
\fi

With all components of the gauge field now expressed in terms of \(\beta_\pm\), we use equation~(\ref{eq:metric}) to determine the corresponding bulk metric. Choosing the group element \(b_\pm(r) = \exp{\left( \pm \dfrac{r}{2l}(L_+ - L_-) \right)}\) and performing appropriate coordinate transformations (see Appendix~\ref{app: BTZ metric} for details), the metric takes the following standard BTZ form:
\begin{equation}\label{eq:BTZmetric}
    ds^2 = - f(r)dt^2 + \frac{dr^2}{f(r)} + r^2 \left( d\phi +\frac{ A_\phi}{r^2} dt \right)^2 
\end{equation}
where
\begin{eqnarray}
    f(r) = \frac{r^2}{l^2} -\frac{2 \pi ^2 l^2 \left(\beta _-^2+\beta _+^2\right)
   }{\beta _-^2 \beta _+^2} + \frac{A_\phi^2}{r^2}, \quad \text{and} \quad A_\phi = \pi ^2  l^3 \left(\frac{1}{\beta _+^2}-\frac{1}{\beta
   _-^2}\right).
\end{eqnarray}

It follows from equation~(\ref{eq:vsigmabetapm}) that the fields \(p_\pm\) are given by  
\begin{equation}\label{eq:p-betarel}
    p_\pm = \pm \frac{2\pi l}{\beta_\pm}.
\end{equation}
Since \(\beta_\pm > 0\), representing the lengths of the Euclidean time circles in the two sectors, the quantity \(p_-\) is always negative for a BTZ black hole.

Finally, the proportionality constants \(\mathcal{C}_\pm\) are determined as
\begin{equation}
    \cC_\pm=\frac{\beta_\pm}{\pi l^2}.
\end{equation}

The solution describes a rotating BTZ black hole with two horizons given by
\begin{equation}
    r_\pm = \pi l^2 \left( \frac{1}{\beta_+} \pm \frac{1}{\beta_-}\right).
\end{equation}

The metric describes a BTZ black hole characterized by its mass \( M \) and angular momentum \( J \), given by  
\begin{equation}\label{eq:MJidentification}
    M = \frac{\pi ^2 \sigma ^2+v^2}{8 G}, \quad \tand \quad J = \frac{\pi  l \sigma  v}{4 G}.
\end{equation}
This relation indicates that a one-dimensional fluid with constant density \(\sigma\) and velocity \(v\) can be mapped to a BTZ black hole, with its mass and angular momentum determined by the expressions above. For the resulting geometry to correspond to a physically acceptable black hole, the inequality \(J \leq lM\) must hold. This requirement imposes a bound on the fluid velocity:
\begin{equation}
	|v| \leq \pi \sigma.
\end{equation}

In this framework, the fluid density \(\sigma\) is always positive, whereas the velocity \(v\) can have either sign. The sign of \(v\) determines the direction of the black hole’s angular momentum. Conversely, given a BTZ black hole characterized by mass \(M\) and angular momentum \(J\), one can associate a boundary fluid with constant density and velocity, determined by
\begin{equation}\label{eq:invMJidentification}
	v \sigma = \frac{4 G J}{\pi  l } \quad \tand \quad \sigma^2 = \frac{4 G \left(\sqrt{l^2
			M^2-J^2}+l M\right)}{\pi ^2 l}.
\end{equation}

The black hole possesses two horizons, denoted by \(r_+\) and \(r_-\), located at
\begin{eqnarray}\label{eq:rpmvalues}
	r_+ = l \pi \sigma \quad \tand \quad r_- = |l v|.
\end{eqnarray}
The extremal black hole limit is reached when \(|v| = \pi \sigma\), while the non-rotating case corresponds to \(v = 0\), which describes a static fluid configuration on the boundary.

The Euclidean temperature of the black hole is determined by requiring the absence of a conical singularity at the outer horizon \(r = r_+\). This leads to the condition
\begin{equation}
\beta = \frac{\beta_+ + \beta_-}{2},
\end{equation}
from which the black hole temperature is found to be
\begin{equation}
    T = \frac{1}{\beta} = \frac{\pi ^2 \sigma ^2-v^2}{2 \pi
   ^2 l \sigma }.
\end{equation}
This result confirms that the triviality of holonomies along the thermal circles in the two sectors ensures the regularity of the Euclidean geometry at the horizon.

Moreover, the difference between \(\beta_+\) and \(\beta_-\) encodes the angular velocity of the black hole, and is given by  
\begin{equation}
 \frac{\beta_- - \beta_+}{2} = \frac{2 \pi  l v}{\pi ^2 \sigma
   ^2-v^2} = \frac{v }{\pi  \sigma } \beta
\end{equation}
where \(\Omega = \dfrac{v}{l \pi  \sigma }\) denotes the angular velocity of the black hole.

\subsection{Relativistic Hamiltonian}\label{sec:relHam}

Among the infinitely many admissible choices for the boundary Hamiltonian, another natural and physically motivated choice is given by
\begin{equation}\label{eq:relHam}
H^\pm_{(2)} = \pm \frac{\mathcal{C}_\pm k}{4\pi}  \int d\theta \left(\frac{p^2_\pm(t,\theta)}{2} + W(\theta) p_\pm(t,\theta) \right).
\end{equation}
As discussed in~\cite{Dutta:2023uxe}, this Hamiltonian admits a phase space interpretation in terms of relativistic fermions on \(S^1\). It also provides a consistent realization of the boundary conditions~\eqref{eq:boundHam} governing the dynamics of boundary gravitons.

As in the ColFT case, the constants \(\mathcal{C}_\pm\) are fixed to ensure that the resulting bulk geometry assumes the standard BTZ form. The full derivation is outlined in Appendix~\ref{app: BTZ metric}. Importantly, the final expressions for the metric, temperature, angular momentum, and \(\beta_\pm\) remain unchanged. In this relativistic case, the proportionality constants are determined to be \(\mathcal{C}_\pm =\pm \dfrac{1}{l}\).

\subsection{The entropy}

Finally, the entropy of the black hole is given by the area of the outer horizon with radius \(r_+\) as
\begin{equation}\label{eq:BTZentropy}
    S = \frac{2\pi r_+}{4G} = \frac{\pi^2 l^2}{2G} \left( \frac{1}{\beta_+} + \frac{1}{\beta_-} \right) .
\end{equation}

Our next objective is to quantize the system and construct the associated Hilbert space \(\mathcal{H}\). With the Hilbert space in hand, we aim to determine the degeneracy of quantum states in \(\mathcal{H}\) that correspond to fixed macroscopic parameters \(M\) and \(J\) (or equivalently, the inverse temperatures \(\beta_\pm\)). This microscopic degeneracy, when appropriately counted, should reproduce the Bekenstein--Hawking entropy, thereby offering a quantum statistical interpretation of the black hole's thermodynamic entropy.

We also formulate the canonical partition function for the black hole and compute the corresponding free energy. Furthermore, we evaluate the logarithmic correction to the Bekenstein--Hawking entropy arising from quantum fluctuations, and demonstrate that this correction is one-loop exact for the class of boundary conditions under consideration.

%%%%%%%%%%%%%%%%%%%%%%%%%%%%%%%%%%%%%%%%%%%%%%%%%%%%%%%%%%%%%%%%%%%%%%%%%%%%%%%%%%%%%%%%%%%%%%%%%%%%%%%%%%%%%%%%%%%%%%%%%%%%%%%%%%%%%%%%%%%%%%

\section{Quantisation and black hole microstates}
\label{sec:quantisation}
As discussed previously, under the boundary conditions under consideration, the spacetime geometry is fully encoded in the two functions \( p_\pm(t,\theta) \), which obey the classical Poisson bracket structure given in equation~(\ref{eq:poisson}). To transition from the classical theory to its quantum counterpart, we now proceed with quantization. This involves promoting the classical Poisson algebra to a quantum operator algebra and constructing an appropriate representation on a Hilbert space.

Since the \(+\) and \(-\) sectors are dynamically independent, the total Hilbert space \(\mathcal{H}\) factorizes into a tensor product of two decoupled sectors:
\[
\mathcal{H} = \mathcal{H}^+ \otimes \mathcal{H}^-.
\]
Our objective is twofold: first, to construct a class of quantum states in \(\mathcal{H}\) such that the expectation value of the metric operator \(d\hat{s}^2\) in these states reproduces the classical spacetime geometry; and second, to determine the degeneracy of such states, thereby providing a microscopic explanation for the entropy of the black hole.

The quantization process begins with a rescaling of the classical fields \( p_\pm(t, \theta) \), defined as
\begin{equation}\label{eq:defptilde}
	p_\pm(t, \theta) = 2\pi\sqrt{\frac{2}{\mathrm{k}}} \tilde p_\pm(t, \theta).
\end{equation}
The new fields \( \tilde p_\pm(t, \theta) \) satisfy the Poisson bracket
\begin{equation}\label{eq:ptildepoisson}
	\{\tilde p_\pm(t, \theta), \tilde p_\pm(t, \theta')\}_{PB} = \mp \frac{1}{2\pi} \frac{\partial}{\partial \theta}\delta(\theta - \theta').
\end{equation}
To quantize the theory, we promote this Poisson bracket to a commutator. Specifically, the right-hand side of equation~(\ref{eq:ptildepoisson}) is multiplied by \(i c^{-1}\), with
\begin{equation}
	c = \frac{3l}{2G},
\end{equation}
where \(c^{-1}\) serves as an effective Planck constant. In the limit \(c \to \infty\), the classical theory is recovered. The parameter \(c\) also plays a crucial role in determining the quantum spectrum and governs the level spacing, directly influencing the count of black hole microstates.

The commutator for the operators \(\tilde{p}_\pm(t,\theta)\) then becomes
\begin{equation}
	[\tilde p_\pm(t,\theta), \tilde p_\pm(t, \theta')] = \mp \frac{i}{2\pi c} \frac{\partial}{\partial \theta}\delta(\theta-\theta').
\end{equation}
To analyze the quantum theory, we expand the fields \(\tilde{p}_\pm(t,\theta)\) into Fourier modes (with implicit time dependence of the mode coefficients \(\alpha_n^\pm\)):
\begin{equation}\label{eq:ptildeexpan}
	\tilde p_\pm(t,\theta) = \frac{1}{2\pi \sqrt{c}} \sum_n \alpha^{\pm}_{\pm n} e^{i n \theta}.
\end{equation}
The mode operators \(\alpha_n^\pm\) satisfy the standard commutation relations of the Kac-Moody algebra:
\begin{equation}\label{eq:balgebra}
	[\alpha^\pm_m , \alpha^\pm_{n}] = m \delta_{m+n}.
\end{equation}

With this operator structure in place, we are now ready to construct the Hilbert space of the theory and identify the quantum states that correspond to various classical \(AdS_3\) geometries.

\subsection{Bosonisation and the Hilbert space}

Given that the modes of \(\tilde{p}_\pm(t, \theta)\) realize two independent copies of the Kac-Moody algebra in equation~(\ref{eq:balgebra}), we employ the bosonization of relativistic Dirac fermions \(\psi_\pm(t, \theta)\) to construct the Hilbert space. This technique establishes a correspondence between these fermionic fields and the bosonic fields \(\tilde{p}_\pm(t, \theta)\)\footnote{See \cite{Afshar:2017okz,Afshar:2016uax,Sheikh-Jabbari:2016npa} for similar constructions.}. To simplify the presentation, we temporarily suppress the \(\pm\) indices, since both sectors are structurally identical. We will reintroduce them when discussing the identification of quantum states corresponding to specific bulk geometries.

We begin with the theory of free Dirac fermions in \( (1+1) \) dimensions. The fermionic fields admit the following mode expansions (suppressing time dependence of the modes):
\begin{eqnarray}\label{eq:fermionmodes}
	\psi(t,\theta) = \frac{1}{\sqrt{2\pi}} \sum_{m \in \mathbb{Z}} \psi_{m-1/2} \ e^{i m \theta}, \quad \psi^\dagger(t,\theta) = \frac{1}{\sqrt{2\pi}} \sum_{m\in \mathbb{Z}} \psi^\dagger_{m-1/2} \ e^{-i m \theta}.
\end{eqnarray}
The fermionic modes \(\psi_q\) and \(\psi^\dagger_r\), with \( q, r \in \mathbb{Z} + \frac{1}{2} \), obey the anti-commutation relations:
\begin{eqnarray}\label{eq:psialgebra}
	\{\psi_q, \psi_r\} = 0, \quad \{\psi^\dagger_q, \psi^\dagger_r\} = 0, \quad \{\psi_q, \psi^\dagger_r\} = \delta_{q,r}.
\end{eqnarray}
These are consistent with the equal-time anti-commutation relations:
\begin{eqnarray}
	\begin{split}
		\{\psi(t,\theta),\psi(t,\phi)\} &= 0, \quad \{\psi^\dagger(t,\theta),\psi^\dagger(t,\phi)\} = 0, \\
		\{\psi(t,\theta),\psi^\dagger(t,\phi)\} &= \delta(\theta-\phi).
	\end{split}
\end{eqnarray}

We define normal ordering by
\begin{eqnarray}
	: \psi^\dagger_q \psi_r : = \begin{cases}
		\ \ \psi^\dagger_q \psi_r & r>0, \\
		- \psi_r \psi^\dagger_q & r<0.
	\end{cases}
\end{eqnarray}
Equivalently, this can be written as
\begin{equation}
	: \psi^\dagger_q \psi_r : = \psi^\dagger_q \psi_r - \langle\psi^\dagger_q \psi_r \rangle.
\end{equation}

We now define a class of bosonic operators following \cite{Avan:1992gm}:
\begin{equation}\label{eq:Boperators}
	B^K_n = \sum_{m\in \mathbb{Z}} m^K : \psi^{\dagger}_{m-n-1/2} \psi_{m-1/2} :.
\end{equation}
These operators satisfy the commutation relation
\begin{eqnarray}\label{eq:BKcommutation}
	\begin{split}
		[B^{K_1}_m, B^{K_2}_n]  = \sum_{q\in \mathbb{Z}} & \left( (q-n)^{K_1}q^{K_2} - (q-m)^{K_2}q^{K_1} \right) : \psi^{\dagger}_{q-m-n-1/2} \psi_{q-1/2} : \\
		& + \delta_{m+n} \sum_{l=0}^{m-1} l^{K_1}(l-m)^{K_2}.
	\end{split}
\end{eqnarray}

The fermion bilinear \(:\psi^\dagger(t,\theta)\psi(t,\theta):\) is then written as
\begin{eqnarray}\label{eq:bosonisation}
	\begin{split}
		:\psi^\dagger(t,\theta) \psi(t,\theta): &= \frac{1}{2\pi} \sum_{m,n \in \mathbb{Z}} :\psi^\dagger_{m-n-1/2} \psi_{m-1/2}: \, e^{i n \theta} = \frac{1}{2\pi} \sum_{n\in \mathbb{Z}} B^0_n e^{i n \theta}.
	\end{split}
\end{eqnarray}
The operators \(B^0_n\) satisfy the Kac-Moody algebra:
\begin{equation}
	[B^0_m, B^0_n] = m \delta_{m+n}.
\end{equation}
We identify \( B^0_n \equiv \alpha_n \) as the modes of the bosonic field \(\tilde{p}(t,\theta)\). Therefore, from equation~(\ref{eq:ptildeexpan}), the bosonization relation takes the form:
\begin{equation}\label{eq:bosonisation2}
	:\psi^\dagger(t,\theta) \psi(t,\theta): \ \equiv \sqrt{c}\, \tilde{p}_\pm(t,\theta).
\end{equation}
This relation shows that the collective fermion bilinear behaves as a bosonic excitation—an essential feature of bosonization\footnote{For similar constructions and reviews, see \cite{Dhar:1992hr,Dhar:2005qh, Das:1995gd, Das:2004rx, Rao:2000rh, Miranda:2003ga}.}.

The inverse relation is given by
\begin{equation}
	\psi(\theta) = \frac{1}{\sqrt{2\pi}} : e^{-2\pi i \varphi(\theta)} :,
\end{equation}
with
\begin{equation}
	\tilde p(\theta) = \frac{ \varphi'(\theta)}{\sqrt{c}}.
\end{equation}

Higher-order fermion bilinears can also be written in terms of higher powers of the bosonized field \(\tilde{p}(\theta)\). For example,
\begin{eqnarray}\label{eq:ptildesquare}
	\begin{split}
		2\pi c :\frac{\tilde p^2(\theta)}{2}: \ & = -\frac{i}{2} :\left( \psi^\dagger \partial_\theta \psi - \partial_\theta  \psi^\dagger \psi\right) : \\
		& =  \frac{1}{2\pi} \sum_p \left(B_p^1 - \frac{p}{2} B_p^0 \right)e^{i p \theta}.
	\end{split}
\end{eqnarray}
Similarly, for the cubic term:
\begin{eqnarray}
	\begin{split}
		(2\pi)^2 c^{3/2} :\frac{\tilde p^3(\theta)}{3}: \ & = -\frac{1}{6} :\partial_{\theta}^2\psi^\dagger\psi + \psi^\dagger \partial_\theta^2\psi - 4 \partial_\theta\psi^\dagger\partial_\theta\psi: \\
		& =  \frac{1}{2\pi} \sum_p \left( B^2_p - p B^1_p + \frac{p^2}{6} B^0_p \right) e^{i p \theta}.
	\end{split}
\end{eqnarray}
Using the bosonization relation~(\ref{eq:ptildesquare}), we find that
\begin{equation}
	:\hat p^2_\pm(t,\theta): = \frac{24}{c^2} \sum_p \left(B^1_p -\frac{p}{2}B_p^0 \right) e^{i p \theta}.
\end{equation}
The boundary Hamiltonian~(\ref{eq:HamCFT}) in this language becomes
\begin{equation}\label{eq:bosonizedH}
	H^\pm = \pm \mathcal{C}_\pm \frac{\sqrt{3}}{c^2} B^2_0.
\end{equation}
Likewise, all conserved charges~(\ref{eq:Qn}) (for \( W(\theta) = 0 \)) take the form
\begin{equation}
	Q^\pm_n = \pm \frac{1}{2\sqrt{3}}\left( \frac{6}{c^2}\right)^n B^{2n}_0.
\end{equation}
Finally, using the algebra~(\ref{eq:BKcommutation}), one confirms that all \( Q^\pm_n \) commute among themselves, affirming the integrable structure of the theory.

%%%%%%%%%%%%%%%%%%%%%%%%%%%%%%%%%%%%%%%%%%%%%%
%%%%%%%%%%%%%%%%%%%%%%%%%%%%%%%%%%%%%%%%%%%%%%
%%%%%%%%%%%%%%%%%%%%%%%%%%%%%%%%%%%%%%%%%%%%%%

\subsection{The Hilbert space}

\begin{figure}[h]
	\centering
	\begin{subfigure}[b]{0.35\textwidth}
		\centering
		\includegraphics[width=0.35\textwidth]{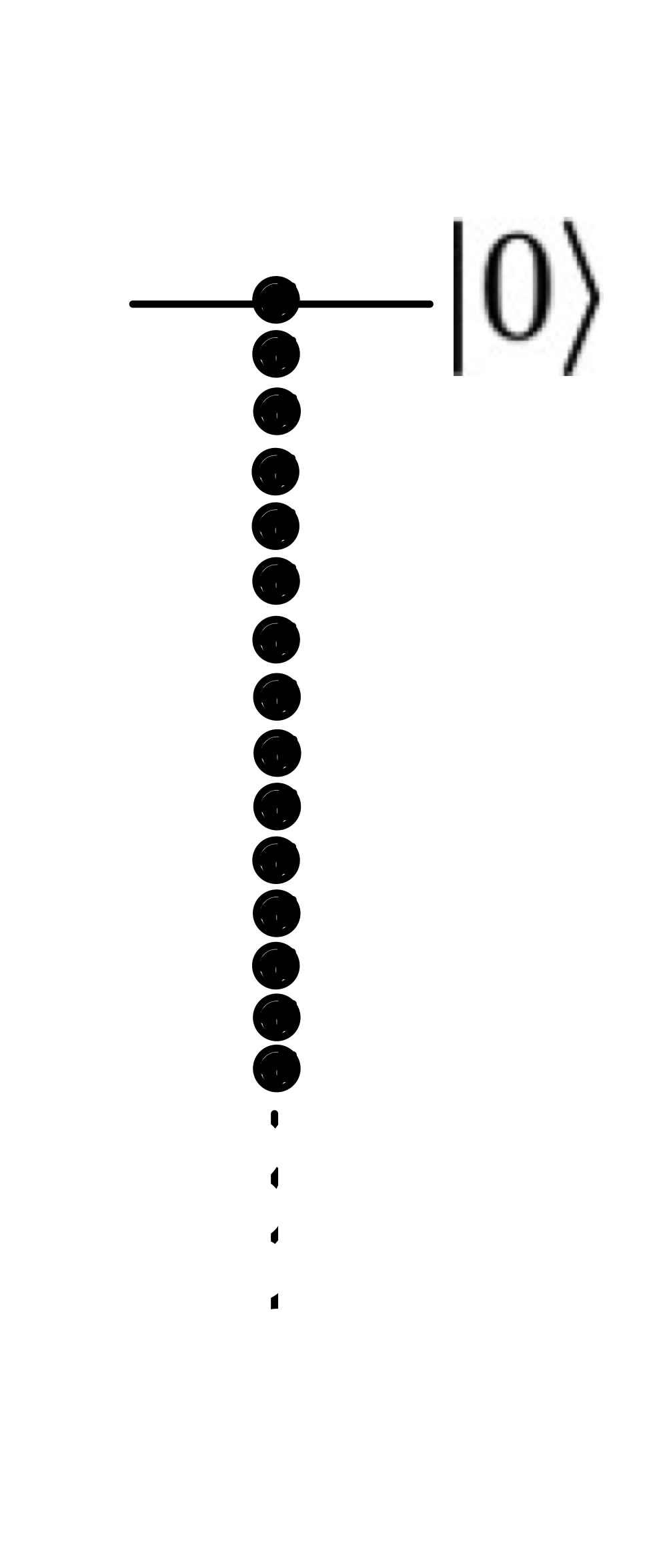}
		\caption{}
		\label{fig:absolutegs}
	\end{subfigure}
	\
	\begin{subfigure}[b]{0.5\textwidth}
		\centering
		\includegraphics[width=0.5\textwidth]{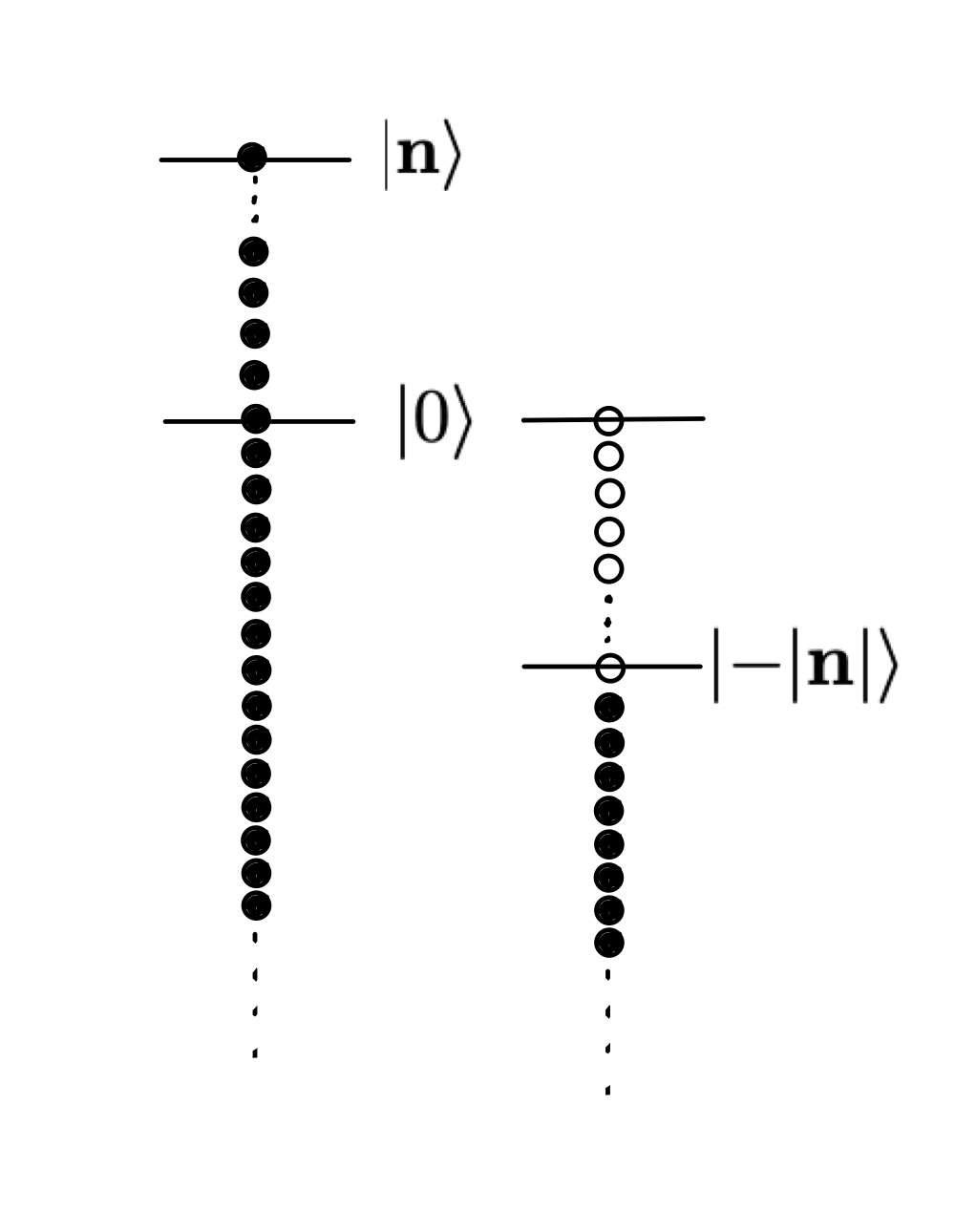}
		\caption{}
		\label{fig:npstate}
	\end{subfigure}
	\caption{(a) Absolute ground state \(\ket{0}\). (b) An \(n\)-particle ground state \(\ket{\mathbf{n}}\) for \(\mathbf{n}>0\) and a hole state for \(\mathbf{n}<0\).}
	\label{fig:gsandnps}
\end{figure}

To construct a representation of the algebra~\eqref{eq:psialgebra}, we begin by defining the Dirac vacuum state \(\ket{0}\), which satisfies
\begin{eqnarray}
	\psi_{k+\frac12}\ket{0} &=& 0, \quad \forall \ k \geq 0, \\
	\psi^\dagger_{k-\frac12}\ket{0} &=& 0, \quad \forall \ k \leq 0.
\end{eqnarray}
This vacuum corresponds to the filled Dirac sea, where all negative energy states up to \(k=0\) are occupied (see Fig.~\ref{fig:absolutegs}). Excitations above this ground state are constructed by acting with creation operators \(\psi^\dagger_q\) for \(q > 0\), and annihilation operators \(\psi_q\) for \(q < 0\). The former represents a particle added above the Dirac sea, while the latter corresponds to removing a fermion from a filled state—i.e., creating a hole.

A general excited state containing \(n_p\) particles and \(n_h\) holes above the Dirac vacuum can be written as
\begin{equation}
	\prod_{i=1}^{n_p} \prod_{j=1}^{n_h} \psi^\dagger_{n_i +\frac12} \psi_{-m_j -\frac12}\ket{0}.
\end{equation}

We define the \(\mathbf{n}\)-particle ground state as
\begin{equation}\label{eq:nstate}
	\ket{\mathbf{n}} =
	\begin{cases}
		\psi^\dagger_{\mathbf{n} - \frac12} \cdots \psi^\dagger_{\frac12}\ket{0} & \mathbf{n} >0, \\
		\psi_{\mathbf{n} + \frac12} \cdots \psi_{-\frac12}\ket{0} & \mathbf{n} < 0.
	\end{cases}
\end{equation}

Using the commutators
\begin{eqnarray}\label{eq:Bpsicommutator}
	[B^K_m, \psi_{k-\frac12}] = - (k+m)^K\psi_{k + m -\frac12}, \quad [B^K_m, \psi^\dagger_{k-\frac12}] = (k-m)^K\psi^\dagger_{k - m -\frac12},
\end{eqnarray}
it follows that
\begin{eqnarray}
	B^0_0 \ket{\mathbf{n}} = \mathbf{n} \ket{\mathbf{n}}, \qquad B^0_m \ket{\mathbf{n}} = 0, \quad \forall \ m>0.
\end{eqnarray}

The Hilbert space \(\mathcal{H}_\mathbf{n}\) consists of all particle-hole excitations built upon the \(\mathbf{n}\)-particle ground state \(\ket{\mathbf{n}}\). As shown in \cite{Haldane:1981zza}, this space can be spanned by bosonic excitations of the form
\begin{equation}\label{eq:kstate}
	\ket{\vec{k}; \mathbf{n}} = \prod_{m>0} \left(B^0_{-m} \right)^{k_m} \ket{\mathbf{n}},
\end{equation}
where \(k_m\in\mathbb{N}\). The bosonic states \(\ket{\vec{k}; \mathbf{n}}\) are eigenstates of \(B^0_0\) with eigenvalue \(\mathbf{n}\), and they obey the orthonormality relation
\begin{equation}\label{eq:knormalisation}
	\braket{\vec{k};\mathbf{n}}{\vec{k}';\mathbf{n}} = z_{\vec{k}} \delta_{\vec{k},\vec{k}'}, \quad z_{\vec{k}} = \prod_m k_m! m^{k_m}.
\end{equation}

Thus, the full fermionic Hilbert space decomposes as
\begin{equation}
	\mathcal{H} = \bigoplus_{\mathbf{n} \in \mathbb{Z}} \mathcal{H}_\mathbf{n}.
\end{equation}

An alternative basis for \(\mathcal{H}_\mathbf{n}\), particularly useful for black hole microstate counting, is given by multi-particle–hole excitations above \(\ket{\mathbf{n}}\). For \(\mathbf{n}>0\), such states take the form
\begin{equation}\label{eq:mnstate}
	\ket{\vec{m}; \vec{n};\mathbf{n}} = \prod_{i} \psi^\dagger_{\mathbf{n}+n_i +\frac12} \psi_{\mathbf{n}- m_i-\frac12}\ket{\mathbf{n}},\quad n_i, m_i \geq 0.
\end{equation}
These states can be encoded in Young diagrams \(\ket{R}\), where the row and column lengths relate to excitation levels via \(n_i = l_i - i\), \(m_i = v_i - i\). The states \(\ket{R;\mathbf{n}} \equiv \ket{\vec{m}; \vec{n};\mathbf{n}}\) are orthonormal:
\[
\braket{R'; \mathbf{n}}{R;\mathbf{n}} = \delta_{RR'}
\]
and can also be written as linear combinations of \(\ket{\vec{k}; \mathbf{n}}\) basis vectors.

For \(\mathbf{n}<0\), we define the excited states analogously:
\begin{equation}\label{eq:mnstate-}
	\ket{\vec{m}; \vec{n};\mathbf{n}} = \ket{R;\mathbf{n}} = \prod_{i} \psi^\dagger_{\mathbf{n} + n_i + \frac12} \psi_{\mathbf{n}-m_i-\frac12}\ket{\mathbf{n}},\quad n_i, m_i \geq 0,
\end{equation}
with the Young diagram now interpreted as the transpose of the one in the \(\mathbf{n}>0\) case.

Before concluding this section, we summarize some important results that will be used in later sections.

Using equation~(\ref{eq:Bpsicommutator}), one can show that the excited states \(\ket{R;\mathbf{n}}\), constructed above the \(\ket{\mathbf{n}}\) ground state, are eigenstates of the operators \(B^K_0\). The corresponding eigenvalues take the form
\begin{equation}
  \mathcal{E}^K = \mathcal{E}_R^K(n_i,m_i) + g_K(\mathbf{n}),
\end{equation}
where \(\mathcal{E}_R^K(n_i,m_i)\) arises from the excitations above the ground state and is given by
\begin{equation}
  \mathcal{E}_R^K(n_i,m_i) = \sum_{i} \left((\mathbf{n} + n_i + 1)^K - (\mathbf{n} - m_i)^K \right),
\end{equation}
and \(g_K(\mathbf{n})\) is the eigenvalue of the ground state \(\ket{\mathbf{n}}\), independent of \(n_i\) and \(m_i\). For \(K = 0, 1, 2\), these quantities are given explicitly by:
\begin{eqnarray}
    \mathcal{E}_R^0 & = & 0, \quad g_0(\mathbf{n}) = \mathbf{n}, \label{eq:E0R} \\
    \mathcal{E}_R^1 & = & \sum_i(n_i+m_i+1), \quad g_1(\mathbf{n}) = \frac{\mathbf{n}(\mathbf{n}+1)}{2}, \label{eq:E1R} \\
    \mathcal{E}_R^2 & = & \sum_i \left( \left(n_i+\frac{1}{2}\right)^2 -  \left(m_i+\frac{1}{2}\right)^2\right) + (2 \mathbf{n} +1) \sum_i (n_i+m_i+1), \label{eq:E2R}\ \\
    \text{and} \quad g_2(\mathbf{n}) &=& \frac{\mathbf{n}(\mathbf{n}+1)(2\mathbf{n}+1)}{3}.
\end{eqnarray}

The quantities \(\mathcal{E}_R^K\) can be expressed in terms of the Young diagram data of the representation \(R\), namely a set of non-increasing, non-negative integers \(\{l_i\}\) that specify the number of boxes in the \(i\)th row of the Young diagram \(R\). First, note that using the relation \(m_i = v_i - i\) and \(n_i = l_i - i\) one can show that~\cite{CordesMooreRamgoolam1995, Cordes:1994fc}
\begin{equation}
\sum_i (n_i + m_i + 1) = \sum_{i=1}^{v_1} l_i = |R| \equiv \text{number of boxes in } R,
\end{equation}
and hence 
\begin{equation}\label{eq:E1RC1R}
\mathcal{E}_R^1 = |R|.
\end{equation}

The calculation of \(\mathcal{E}_R^2\) can be carried out separately for the cases \(\mathbf{n} > 0\) and \(\mathbf{n} < 0\). For \(\mathbf{n} > 0\), using the relations between \(n_i, m_i\) and the Young diagram data \(l_i, v_i\), we obtain~\cite{CordesMooreRamgoolam1995, Cordes:1994fc} :
\begin{equation}
    \sum_i \left( \left(n_i+\frac{1}{2}\right)^2 -  \left(m_i+\frac{1}{2}\right)^2\right) = \sum_{i=1}^{v_1} l_i(l_i - 2i) + |R| \equiv \kappa_R.
\end{equation}
Therefore,
\begin{equation}\label{eq:E2RC2Rp}
    \mathcal{E}^2_R = \kappa_R + (2 \mathbf{n}+1) |R|.
\end{equation}

A similar construction applies for \(\mathbf{n} < 0\), where the roles of \(l_i\) and \(v_i\) are reversed, with \(n_i = v_i - i\) and \(m_i = l_i - i\). This corresponds to taking the transpose of the Young diagram used in the \(\mathbf{n} > 0\) case. In this case, we obtain
\begin{equation}\label{eq:E2RC2Rm}
    \mathcal{E}^2_R= - \kappa_R - (2 |\mathbf{n}|-1) |R|, \quad \text{for} \ \mathbf{n} < 0.
\end{equation}

\subsection{States in the Hilbert space and $AdS_3$ geometry}

In the quantum framework, the classical fields \(\sigma(t, \theta)\) and \(v(t, \theta)\) (or equivalently, \(p_\pm(t, \theta)\)) are promoted to operators acting on a Hilbert space. A classical configuration, as described by equation (\ref{eq:BTZmetric}), corresponds to a quantum state \(\ket{\Psi} = \ket{\Psi^+} \otimes \ket{\Psi^-}\) in the Hilbert space \(\mathcal{H} = \mathcal{H}^+ \otimes \mathcal{H}^-\), such that
\begin{equation}
    \bra{\Psi}:\hat v(t,\theta) \hat \sigma(t,\theta):\ket{\Psi} = \frac{4GJ}{\pi l} \quad \text{and} \quad \bra{\Psi}:\pi^2 \hat \sigma^2(t,\theta) + \hat v^2(t,\theta):\ket{\Psi} = 8GM.
\end{equation}
This is equivalently expressed as
\begin{equation}\label{eq:pMJrelation}
    \bra{\Psi^\pm} : \hat p_\pm^2(t,\theta) : \ket{\Psi^\pm} = \frac{12}{c}\left( M l \pm J \right) = \frac{4 \pi ^2 l^2}{\beta_\pm^2}.
\end{equation}

Given a state \(\ket{\Psi^\pm} \in \mathcal{H}^\pm\), one can reconstruct a unique bulk geometry. However, the converse does not necessarily hold: a single bulk geometry may correspond to multiple quantum states, leading to degeneracy.

We now turn to the identification of quantum states in \(\mathcal{H}^\pm\) with their corresponding bulk geometries. For the \(\mathbf{n}_\pm\)-particle ground state, the expectation values of the operator \(\hat{p}_\pm\) and its square are given by (using the relations~(\ref{eq:E0R}) and (\ref{eq:E1R}) with \(|R|=0\))
\begin{equation}
    \bra{\mathbf{n}_\pm}\hat{p}_\pm \ket{\mathbf{n}_\pm} = \frac{2\sqrt{3}}{c}\mathbf{n}_\pm, \quad \bra{\mathbf{n}_\pm}\hat{p}^2_\pm \ket{\mathbf{n}_\pm} = \frac{12}{c^2} (\mathbf{n}_\pm)^2.
\end{equation}
Therefore the state \(\ket{\mathbf{n}_+} \otimes \ket{\mathbf{n}_-}\) corresponds to a BTZ black hole geometry, with mass and angular momentum determined by the quantum numbers \(\mathbf{n}_\pm\). In particular, the ground state with \(\mathbf{n}_\pm = 0\) describes a massless BTZ black hole. An interesting feature of the \(\mathbf{n}_\pm\)-particle ground state is that the quantum dispersion of the \(\hat{p}_\pm\) operator, given by \(\langle \hat{p}_\pm^2 \rangle - \langle \hat{p}_\pm \rangle^2\), vanishes identically.

Next, we consider excited states of the form \( \ket{R^\pm;\mathbf{n}_\pm} \). Using the relations~(\ref{eq:E0R}) and (\ref{eq:E1R}), we find
\begin{equation}
    \bra{R^\pm;\mathbf{n}_\pm}B^0_0\ket{R^\pm;\mathbf{n}_\pm} = \mathbf{n}_\pm , \quad \bra{R^\pm;\mathbf{n}_\pm}B^1_0\ket{R^\pm;\mathbf{n}_\pm} = \frac{\mathbf{n}_\pm(\mathbf{n}_\pm + 1)}{2}  + |R^\pm| ,
\end{equation}
The expectation value of the normal-ordered operators \(\hat p_\pm\)  and \(\hat{p}^2_\pm(t,\theta)\) in the excited state are
\begin{equation}\label{eq:expvalueinRn}
    \bra{R^\pm;\mathbf{n}_\pm}\hat{p}_\pm \ket{R^\pm;\mathbf{n}_\pm} = \frac{2\sqrt{3}}{c}\mathbf{n}_\pm, \quad \bra{R^\pm;\mathbf{n}_\pm}\hat{p}^2_\pm \ket{R^\pm;\mathbf{n}_\pm} = \frac{24}{c^2} \left( \frac{\mathbf{n}_\pm(\mathbf{n}_\pm + 1)}{2} +|R^\pm| \right).
\end{equation}
These excited states also describe BTZ black holes with excitations above the background specified by the ground state. We treat this \(\mathbf{n}_\pm\)-particle ground state as our reference configuration or background geometry.

It is important to note that thermal \(AdS\) geometries require \(\langle \hat{p}^2_\pm \rangle < 0\), which in turn implies that the eigenvalues of \(\hat{p}_\pm\) must be imaginary. While it is possible to realize such configurations by constructing complex representations of the algebra---thereby associating a quantum state to thermal \(AdS\) spacetime \cite{Afshar:2017okz}. In in our analysis—both for the microcanonical and canonical ensembles—we choose a reference state \(\ket{\mathbf{n}_\pm}\) and compute the degeneracy and free energy of BTZ black holes relative to this chosen background.

\subsection{Microcanonical ensemble and degeneracy}

To compute the degeneracy of BTZ black holes above a classical background, we identify the quantum states in the Hilbert space that correspond to fixed mass and angular momentum relative to this background geometry, and count the number of such states. Although not strictly necessary, we simplify the analysis by working in the zero-particle sector, i.e., setting \(\mathbf{n}_\pm = 0\). This choice effectively means that we are counting the degeneracy of black hole microstates with respect to the massless BTZ black hole.
Denoting the excited states above the zero-particle ground states by \(\ket{R^\pm}\), we have
\begin{equation}\label{eq:pRrelation}
    \bra{R^\pm}:\hat p^2_\pm(t,\theta): \ket{R^\pm} = \frac{24}{c^2} |R^\pm|.
\end{equation}

By comparing equations (\ref{eq:pMJrelation}) and (\ref{eq:pRrelation}), we find that
\begin{equation}\label{eq:boxnumberMJrel}
    |R^\pm| = \frac{c}{2} \left( M l \pm J \right) = \frac{c^2 \pi ^2 l^2}{6 \beta_\pm^2}.
\end{equation}
The Young diagram states \(\ket{R^\pm}\), each consisting of \(|R^\pm|\) boxes, represent the microstates of a BTZ black hole with mass \(M\) and angular momentum \(J\). Consequently, the degeneracy is given by the number of possible Young diagrams with a fixed total number of boxes \(|R^\pm|\). In the classical limit—corresponding to large central charge \(c\)—the number of boxes becomes large, and the degeneracy can be approximated using the Hardy–Ramanujan formula:
\begin{eqnarray}
    d(p_+, p_-) = e^{2\pi \sqrt{\frac{|R^+|}{6}} + 2\pi \sqrt{\frac{|R^-|}{6}}}.
\end{eqnarray}

This yields the statistical entropy
\begin{equation}
    S_{\text{stat}} = \ln d(p_+, p_-) = 2\pi \sqrt{\frac{|R^+|}{6}} + 2\pi \sqrt{\frac{|R^-|}{6}} = \frac{\pi^2 l^2}{2G} \left( \frac{1}{\beta_+} + \frac{1}{\beta_-} \right),
\end{equation}
which agrees precisely with the thermodynamic entropy given in equation~(\ref{eq:BTZentropy}).

\section{Canonical partition function}
\label{sec:canonical}

In the canonical ensemble, the entropy of a black hole is obtained from the thermodynamic relation
\[
S = \log \mathcal{Z} + \beta \bar{E},
\]
where \(\mathcal{Z}\) is the partition function, \(\beta\) is the inverse temperature, and \(\bar{E}\) is the average energy of the system. At leading order, this reproduces the Bekenstein--Hawking area law, \(S = \frac{A}{4G},\) where \(A\) is the horizon area. Subleading corrections to this result arise from quantum and thermal fluctuations around the classical background. These typically include logarithmic terms of the form \(\alpha \log A\), where the coefficient \(\alpha\) depends on the field content (see \cite{Sen:2012dw} for details), boundary conditions, choice of ensemble, and the quantum gravity framework under consideration. It was proposed in~\cite{Carlip:2000nv} that if the black hole entropy arises from a single conformal field theory, and if the central charge \(c\) is \emph{universal}—that is, independent of the horizon area—then the logarithmic correction coefficient \(\alpha\) universally takes the value \(-\tfrac{3}{2}\). Logarithmic corrections provide important insights into the microscopic structure of black hole entropy and serve as consistency checks for candidate theories of quantum gravity.

Our objective is to compute the coefficient \(\alpha\) associated with the logarithmic correction to black hole entropy under Kac--Moody boundary conditions, for two distinct choices of boundary Hamiltonians: the ColFT Hamiltonian~\eqref{eq:HamCFT} and the relativistic Hamiltonian~\eqref{eq:relHam}. Remarkably, in both cases we find
\[
\alpha = -\frac{1}{2}.
\]
A similar value for the logarithmic correction coefficient was also obtained in~\cite{Mukherji:2002de} in the context of \(AdS_5\) black holes.

To obtain this result, we employ a novel method to compute the canonical partition function up to subleading order. Specifically, we demonstrate that the partition function in each sector resembles that of two-dimensional chiral \(U(N)\) Yang--Mills theory on a torus. Taking the classical limit of our setup corresponds to the large-\(N\) limit of Yang--Mills theory, allowing us to systematically extract both the leading and subleading contributions. We first present the computation for the ColFT Hamiltonian, followed by the relativistic case.

\subsection{Collective field theory Hamiltonian}\label{eq:CanoCFT}

Euclidean \(AdS_3\) gravity can be reformulated as two decoupled \(SL(2,\mathbb{R})\) Chern--Simons theories, each defined on a solid torus. The periodicities of the Euclidean time direction in the two chiral sectors are denoted by \(\beta_\pm\). The dynamics of the theory is determined by the specific choice of boundary Hamiltonians. The canonical partition function \(\mathcal{Z} \sim e^{-\beta M + \beta \Omega J}\) encodes the classical energy, angular momentum, and entropy of the black hole. At the classical level, the combination \(-\beta M + \beta \Omega J\) can be shown to coincide with \(- \beta_+ H^+_{(1)} - \beta_- H^-_{(1)}\), where \(H^\pm_{(1)}\) is defined using the Hamiltonian~(\ref{eq:HamCFT}) with \(W(\theta) = 0\). This equivalence can be explicitly verified by substituting the classical values of the fields \(p_\pm\) into the Hamiltonian expression. As a result, the full canonical partition function of the BTZ black hole, evaluated at fixed \(\beta_\pm\), takes the form:
\begin{equation}\label{eq:canoZd}
\mathcal Z = \Tr \exp\left({-\beta_+ H^+_{(1)}}\right) \Tr \exp\left(-\beta_- H^-_{(1)}\right)
\end{equation}
where the trace is taken over the Hilbert space \(\mathcal{H}^\pm\) that keeps the inverse temperature \(\beta_\pm\) fixed.

Our objective is to compute the free energy \(\mathcal{F} = \ln \mathcal{Z}\) with respect to the background geometry defined by the \(\mathbf{n}_\pm\)-particle states. Unlike the microcanonical case, we cannot set \(\mathbf{n}_\pm = 0\) here. The choice of \(\mathbf{n}_\pm\) is dictated by the requirement that the black hole geometry and the background geometry share the same Euclidean structure. Since the size of the Euclidean time circle is determined by the expectation value of \(\hat{p}_\pm\), as given in equation~(\ref{eq:p-betarel}), we obtain
\begin{equation}
    \mathbf{n}_\pm = \pm \frac{\pi c l}{\sqrt{3}\beta_\pm}.
\end{equation}

In the canonical ensemble, all excited states \(\ket{R^\pm;\mathbf{n}_\pm}\) preserve the expectation value of \(\hat{p}_\pm\), and therefore contribute to the partition function~(\ref{eq:canoZd}). Thus the partition function is give by, 
\begin{equation}\label{eq:canoZ1}
    \mathcal{Z} = \sum_{R^+} \exp\left[-\frac{\sqrt{3}}{\pi c^2}\left(\frac{\beta_+}{l}\right)^2 \bra{R^+;\mathbf{n}_+} B^2_0 \ket{R^+;\mathbf{n}_+}\right] \cdot \sum_{R^-} \exp\left[\frac{\sqrt{3}}{\pi c^2}\left(\frac{\beta_-}{l}\right)^2 \bra{R^-;\mathbf{n}_-} B^2_0 \ket{R^-;\mathbf{n}_-}\right].
\end{equation}
The sum runs over all possible representations. From (\ref{eq:E2RC2Rp}) and (\ref{eq:E2RC2Rm}) we find that the expectation value of \(B^2_p\) over the states \(\ket{R^\pm;\textbf{n}_\pm}\) is given by
\begin{eqnarray}\label{eq:B2exp}
\begin{split}
    \bra{R^\pm;\mathbf{n}_\pm}B^2_0\ket{R^\pm;\mathbf{n}_\pm} & = \pm (\kappa_{R^\pm} + 2 |\mathbf{n}_\pm| |R^\pm|) + g_2(\mathbf{n}_\pm)\\
    & = \pm \left(\sum_i l^\pm_i(l^\pm_i - 2i) + |R^\pm| + (2 |\mathbf{n}_\pm|\pm 1) |R^\pm|\right) + g_2(\mathbf{n}_\pm)
    \end{split}
\end{eqnarray}

Since we are interested in computing the free energy relative to the background geometry, we subtract the energy of the background configuration from the Hamiltonians appearing in equation~(\ref{eq:canoZd}). Accordingly, we omit the last term \(g_2(\mathbf{n}_\pm)\) in~(\ref{eq:B2exp}), as it corresponds to the energy of the background geometry.

Accordingly, the canonical partition functions in the \(+\) and \(-\) sectors take the form
\begin{equation}\label{eq:nonrelPF}
    \mathcal{Z}_\pm = \sum_{R^\pm} \exp\left[- \frac{\sqrt{3}}{\pi c^2} \left(\frac{\beta _\pm}{l}\right)^2  \left(\sum_i l^\pm_i(l^\pm_i +1 - 2i) + (2 |\mathbf{n}_\pm|\pm 1) |R^\pm|\right)\right].
\end{equation}

In general, the right-hand side of equation~(\ref{eq:nonrelPF}) is difficult to evaluate exactly. However, one can extract both the leading and subleading contributions to the free energy in the classical limit.

To facilitate further analysis, we rearrange the terms in the exponential and express the partition function in the following form\footnote{Here we have assumed that \(|\mathbf{n}_\pm| \gg 1\).}:
\begin{equation}
    \mathcal{Z_\pm} = 
    \sum_{R^\pm} \exp\left[- \frac{\tilde A_\pm}{2 |\mathbf{n}_\pm|}\sum_i l^\pm_i(l^\pm_i +1 - 2i) - \tilde A_\pm \sum_i l^\pm_i\right]
\end{equation}
where,
\begin{equation}
    \tilde A_\pm = 2 \left(\frac{\beta_\pm}{c\, l}\right).
    %\qquad \text{and} \quad N_\pm = |\mathbf{n}_\pm| = \frac{2 \pi\,  c \, l}{\sqrt{3}\beta_\pm}.
\end{equation}
This expression closely resembles the partition function of chiral \(U(N)\) Yang--Mills theory on a torus\footnote{The summation in the exponential runs over all \(1 \leq i < \infty\). To regulate the expression, we introduce a cutoff by restricting the sum to \(1 \leq i < |\mathbf{n}_\pm|\), and eventually take the limit \(|\mathbf{n}_\pm| \rightarrow \infty\).
} with \(N_\pm = |\mathbf{n}_\pm|\), where \(\tilde A_\pm\) plays the role of the torus area \cite{Gross:1992tu, Gross:1993hu,Gross:1993yt, Migdal1975, Rusakov1990, Chattopadhyay:2020rle, Cordes:1994sd, Ramgoolam:1993hh}. 

One can compute the free energy in each chiral sector from the corresponding partition function via the standard relation
\begin{equation}
    \mathcal{F}^\pm = \ln \mathcal{Z}_\pm.
\end{equation}
This free energy admits a genus expansion of the form
\begin{equation}
    \mathcal{F}^\pm = \sum_{g=0}^\infty N_\pm^{2-2g} F^\pm_g(\tilde A_\pm).
\end{equation}
The leading and next-to-leading contributions are given explicitly by~\cite{Okuyama:2018clk, Rudd:1994ta, Douglas:1993wy}
\begin{equation}
    F^\pm_0(\tilde A_\pm) = -\frac{\tilde A_\pm}{48}, \quad F^\pm_1(\tilde A_\pm) = - \log \eta(Q_\pm),
\end{equation}
where \(\eta(Q_\pm) := Q_\pm^{1/24} \prod_{n=1}^\infty (1 - Q_\pm^n)\) is the Dedekind eta function, and \(Q_\pm = \exp[-\tilde A_\pm/2]\).

Higher-genus terms \(F_g(\tilde A_\pm)\) for \(g \geq 2\) can also be systematically computed; closed-form expressions and recursive techniques are provided in~\cite{Rudd:1994ta}. In this expansion, the area \(\tilde A_\pm\) is kept fixed while the series is developed in inverse powers of \(N_\pm\). At first glance, one might expect that the classical limit \(cl/\beta_\pm \to \infty\) is dominated solely by the genus-zero contribution \(F_0\). However, in this limit, the area \(\tilde A_\pm\) also tends to zero, and the contributions from higher genera become significant.

To account for this, one needs to analyze the small-\(\tilde A_\pm\) behavior of the genus-\(g\) free energies. As shown in~\cite{Rudd:1994ta}, this behavior is given by
\begin{equation}
    F_g(\tilde A_\pm) \sim (\tilde A_\pm)^{-2g+1} \pi^{2g} \mathbf{c}_g, \quad g \geq 2,
\end{equation}
where \(\mathbf{c}_g\) are rational numbers. Since each \(F_g\) term appears at order \(N_\pm^{2 - 2g}\), and \(N_\pm \sim 1/\tilde{A}_\pm\), the contribution from genus \(g\) to the total free energy \(\mathcal{F}^\pm\) scales as \(\sim \pi^{2g} \mathbf{c}_g / \tilde{A}_\pm\). Hence, the contributions from genus two and higher scale similarly to the genus-zero term: \(\sim \tilde{A}_\pm N_\pm^2 \sim 1/\tilde{A}_\pm\). Consequently, in the classical limit where \(\tilde{A}_\pm \to 0\), contributions from all genera become equally significant. This implies that, for an accurate computation of the free energy, one must retain contributions from all genera.

We now evaluate \(F_1(\tilde A_\pm)\) in the small \(\tilde A_\pm\) limit (see Appendix~\ref{app:Infinite product computation}). In this regime, it behaves as
\begin{equation}
    F_1(\tilde A_\pm) = \frac{\pi^2}{3\tilde A_\pm} + \frac{1}{2}\ln \tilde A_\pm.
\end{equation}

Thus, in the limit \(\tilde A_\pm \rightarrow 0\), the dominant contribution to the free energy takes the form
\begin{eqnarray}
    \mathcal{F}^\pm = \mathcal{F}_0^\pm + \frac{1}{2} \ln \tilde A_\pm,
\end{eqnarray}
where
\begin{equation}
    \mathcal{F}_0^\pm =  \frac{\pi^2}{3\tilde A_\pm} \sum_{g\geq 0} \frac{3^g \mathbf{c}_g}{16^{g-1}}, \quad \text{with} \quad \mathbf{c}_0 = -\frac{1}{192}, \ \text{and} \  \mathbf{c}_1 = \frac{1}{3}.
\end{equation}
The dominant free energy can be written as
\begin{equation}\label{eq:domF-colft}
    \mathcal{F}_0^\pm = \frac{\pi^2}{3 \tilde A_\pm} - \frac{\pi^2}{3 \tilde A_\pm}\left(\frac{1}{12} - 16\sum_{g \geq 2} \left(\frac{3}{16}\right)^g \mathbf{c}_g \right).
\end{equation}
The first term \(\frac{\pi^2}{3 \tilde A_\pm}\) in \(\mathcal{F}_0^\pm\) reproduces the expected free energy of the BTZ black hole and originates from the genus-one contribution \(F_1(\tilde A_\pm)\). However, \(\mathcal{F}_0^\pm\) also receives additional contributions from the genus-zero term and from all higher-genus terms with \(g \geq 2\). To determine the remaining part of \(\mathcal{F}_0^\pm\), one must perform an infinite sum over these higher-genus corrections.

Carrying out this computation requires knowledge of the exact form of the coefficients \(\mathbf{c}_g\) for all \(g\), which is generally unavailable. A few of these coefficients are known and can be found in~\cite{Rudd:1994ta}, but the full structure remains elusive. Moreover, convergence of the infinite sum is not guaranteed.

The higher-genus contributions \(F_g\) are expressed in terms of Eisenstein series \(E_2\) and their derivatives~\cite{Rudd:1994ta,Okuyama:2019rqn}. A potentially more systematic approach to understanding the convergence and evaluating the full sum would be to exploit the modular properties of these functions to determine \(\mathbf{c}_g\). However, such an analysis is beyond the scope of this work and is left for future investigation.

If the contribution turns out to be non-zero, it would indicate that even in the classical limit, the free energy receives corrections from additional saddle points—arising from the specific choice of boundary Hamiltonian—thereby motivating further investigation. In the next section, we demonstrate that such contributions from additional saddles are absent when the boundary Hamiltonian is chosen to be the phase space Hamiltonian of relativistic fermions.

Importantly, the logarithmic correction to the free energy arises solely from the one-loop contribution, namely \(F_1(\tilde A_\pm)\). Consequently, the correction to the Bekenstein--Hawking entropy for an asymptotically \(AdS_3\) BTZ black hole---whose asymptotic symmetry is governed by a Kac--Moody algebra and whose dynamics are dictated by the ColFT Hamiltonian is given by:
\begin{equation}\label{eq:logcor}
    \Delta S =  \frac{1}{2} \ln \left(\frac{\beta_+ \beta_-}{c^2l^2}\right) = -  \frac{1}{2} \ln \left(\frac{9l^4}{4G^2 \beta^2(1 - l^2 \Omega^2)}\right).
\end{equation}

\subsection{Sub-leading correction to entropy for relativistic Hamiltonian}

We now turn to the computation of the canonical partition function associated with the Hamiltonian defined in equation~(\ref{eq:relHam}). At the classical level, one can verify that the thermodynamic combination \(-\beta M + \beta \Omega J\) corresponds to \(-\beta_+ H^+_{(2)} - \beta_- H^-_{(2)}\). Consequently, the canonical partition function in each chiral sector is given by
\begin{equation}
    \mathcal{Z}_\pm = \mathrm{Tr} \left[\exp\left(- \frac{k}{4\pi} \frac{\beta_\pm}{l} \int d\theta\,\frac{p^2_\pm(t,\theta)}{2}\right)\right] = \mathrm{Tr} \exp\left(-\frac{\beta_\pm}{c\,l} B^1_0(p_\pm)\right).
\end{equation}

In the Young diagram basis \(\ket{R^\pm}\), the operator \(B^1_0(p_\pm)\) acts diagonally,
\begin{equation}
    B^1_0\ket{R^\pm,\textbf{n}}= \left(|R^\pm| +\frac{1}{2} \textbf{n} (\textbf{n}+1)\right)\ket{R^\pm,\textbf{n}}.
\end{equation}
Subtracting the energy of the background geometry we find
\begin{equation}
    \mathcal{Z}_\pm = \sum_{R^\pm} \exp\left(-\frac{\beta_\pm}{c\,l} |R^\pm|\right),
\end{equation}
where \(|R^\pm|\) denotes the total number of boxes in the Young diagram associated with the representation \(R^\pm\), which is also the size of the corresponding integer partition. Letting \(n = |R^\pm|\), and denoting the number of partitions of \(n\) by \(p(n)\), the partition function can be rewritten as
\begin{equation}
     \mathcal{Z}_\pm = \sum_{n=0}^{\infty} p(n) e^{-\mu_\pm n},
\end{equation}
with \(\mu_\pm = \frac{\beta_\pm}{c\,l}\). This expression is the generating function for the integer partition function and is given explicitly by
\begin{equation}
  \mathcal{Z}_\pm = \prod_{k=1}^{\infty} \frac{1}{1 - e^{-\mu_\pm k}}, \quad \mu_\pm > 0.
\end{equation}

In the classical limit \(\mu_\pm \rightarrow 0\), the partition function admits the following asymptotic expansion:
\begin{align}
     \log \mathcal{Z}_\pm = \frac{\pi^2}{6 \mu_\pm} + \frac{1}{2} \log \mu_\pm + \cdots.
\end{align}
The free energy in each chiral sector is given by the logarithm of the corresponding partition function \(\mathcal{F}^\pm = \log \mathcal{Z}_\pm\). Summing over both sectors, we obtain the leading-order contribution to the total free energy of the BTZ black hole:
\begin{equation}\label{eq:FclH1}
 \mathcal{F}_{0} = \frac{l^2 \pi^2}{2 \beta G (1 - l^2 \Omega^2)}.
\end{equation}
Thus, we see that for this choice of Hamiltonian, the free energy precisely reproduces that of the BTZ black hole, in contrast to the earlier case~(\ref{eq:domF-colft}).

From this free energy, one recovers the classical expressions for the BTZ black hole's entropy, mass, and angular momentum. The subleading correction to the entropy, arising from the subleading term in \(\log \mathcal{Z}_\pm\), is given by
\begin{equation}
    \Delta S = -\frac{1}{2} \log\left(\frac{9 l^4}{4 G^2 \beta^2 (1 - l^2 \Omega^2)}\right),
\end{equation}
which precisely matches equation (\ref{eq:logcor}).

\section{Conclusion}\label{sec:conclusion}

In this work, we have investigated three-dimensional gravity in asymptotically \(\mathrm{AdS}_3\) spacetime under a new class of non-standard boundary conditions determined by a dynamical boundary Hamiltonian. Specifically, we focused on the collective field theory (ColFT) Hamiltonian, which governs the dynamics of a one-dimensional compressible fluid on the spatial boundary. These boundary dynamics encode the bulk physics, including the BTZ black hole geometry, and different fluid configurations correspond to different bulk solutions—such as rotating, extremal, and static BTZ black holes.

We carried out a canonical quantization of this system by promoting the classical Poisson algebra to quantum commutators and identifying the resulting quantum theory with a free relativistic fermionic system on a circle. Through bosonization, we constructed the corresponding Hilbert space and established a one-to-one map between bulk gravitational configurations and particle–hole excitations in the fermionic theory. These excitations are labeled by Young diagrams, and we showed that their degeneracy reproduces the Bekenstein–Hawking entropy, providing a microscopic origin of black hole thermodynamics in this setup.

Subsequently, we analyzed the canonical partition function in the Euclidean path integral formulation for two distinct choices of boundary Hamiltonians: the ColFT Hamiltonian and a Hamiltonian for relativistic fermions. For the ColFT case, we found that the partition function factorizes into two chiral sectors, each resembling the partition function of two-dimensional \(U(N)\) Yang–Mills theory on a torus, with the rank scaling as \(N \sim c/(\beta l)\). We computed the leading and subleading contributions to the free energy in the classical limit. Notably, the leading entropy receives contributions from all genera, while the subleading correction arises entirely from the genus-one term and is thus one-loop exact. The coefficient of this logarithmic correction was found to be \(-\frac{1}{2}\). A similar analysis was performed for the relativistic fermion Hamiltonian, where the partition function reduces to the generating function for integer partitions. The same one-loop correction with coefficient \(-\frac{1}{2}\) emerged, indicating the universality of this result across different boundary Hamiltonians.

Our results establish a concrete realization of the holographic principle in three dimensions, where a purely boundary quantum system encodes the thermodynamic and microscopic features of the BTZ black hole. The framework we present opens up several avenues for future investigation. It would be interesting to explore the modular properties of the genus expansion in more detail and examine whether a resummation of the partition function can be performed using techniques from topological string theory. Another direction is to analyze more general boundary Hamiltonians, including those corresponding to interacting fermionic systems or those arising from integrable models. Extensions to higher spin theories, or supersymmetric setups could provide further insights into the understanding of black hole microstate counting.

\vspace{1cm}
\paragraph{Acknowledgment} We thank Nabamita Banerjee, Ranveer Singh for useful discussions. We are indebted to people of India for their unconditional support toward the researches in basic science.

\appendix

\section{BTZ metric}\label{app: BTZ metric}
In this section, we compute the entropy of BTZ black hole from the Bekenstein Hawking area law.

\subsection{Non-relativistic Hamiltonian}
The chemical potentials $\xi^\pm$ corresponding to the rescaled Hamiltonian \eqref{eq:HamCFT} with potential $W(\theta)=0$ can be computed using \eqref{eq:xidef},
\begin{equation}
      \xi^\pm = \mp \cC_\pm \frac{p_\pm^2}{2} .
\end{equation}
Metric obtained from equation \eqref{eq:metric} by choosing gauge connections of the form \eqref{eq:Apm-apm} and \eqref{eq:apmform} is given by,
\begin{align}
    ds^2&=\frac{\pi ^4 \, l^6}{\beta_-^4 \beta_+^4}  \left(\beta_-^4 \cC_+^2+2 \beta_-^2 \beta_+^2 \cC_+ \cC_- \cosh \left(\frac{2 \rho }{l}\right)+\beta_+^4 \cC_-^2\right)dt^2+d\rho^2\nn\\
    &-\frac{2\pi ^3  l^5 }{\beta_-^3 \beta_+^3}\left(\beta_-^3 \cC_+-\beta_- \beta_+ \cosh \left(\frac{2 \rho }{l}\right) (\beta_- \cC_+-\beta_+ \cC_-)-\beta_+^3 \cC_-\right)dt\,d\theta \nn\\
    &+\frac{\pi ^2  l^4 }{\beta_-^2 \beta_+^2}\left(\beta_-^2+\beta_+^2-2 \beta_- \beta_+ \cosh \left(\frac{2 \rho }{l}\right)\right) d\theta^2.
\end{align}

In order to obtain this form, we have taken $p_\pm(t,\theta) = v\pm \pi\sigma=\pm\frac{2 \pi l}{\beta_\pm.} $. We perform a coordinate transformation in the radial coordinate to obtain a simpler form of metric,
\begin{equation}
   \rho=\frac{l}{2} \cosh ^{-1}\left(\frac{\pi ^2\,l^4 (\beta_-^2 + \beta_+^2 )-\beta_-^2 \beta_+^2 r^2}{2 \pi ^2 \beta_- \beta_+ l^4}\right).
\end{equation}
and obtain,
\begin{align}
    ds^2&=\frac{\pi ^2  l^2 }{\beta_-^4 \beta_+^4}\left(\pi ^2 \beta_-^4 \cC_+^2 l^4+\beta_- \beta_+ \cC_+ \cC_- \left(\pi ^2 l^4 \left(\beta_-^2+\beta_+^2\right)-\beta_-^2 \beta_+^2 r^2\right)+\pi ^2 \beta_+^4 \cC_-^2 l^4\right) dt^2\nn\\
    &+r^2\,d\theta^2 + \left(\frac{l^2 r^2\,\beta_-^4 \beta_+^4 }{\pi ^4 l^8 \left(\beta_-^2-\beta_+^2\right)^2-2 \pi ^2 \beta_-^2 \beta_+^2 l^4 r^2 \left(\beta_-^2+\beta_+^2\right)+\beta_-^4 \beta_+^4 r^4} \right)dr^2\nn\\
    &-\frac{\pi  l }{ \beta_-^3 \beta_+^3}\left(\cC_+ \left(\pi ^2 \beta_- l^4 \left(\beta_-^2-\beta_+^2\right)+\beta_-^3 \beta_+^2 r^2\right)-\beta_+ \cC_- \left(\pi ^2 l^4 \left(\beta_+^2-\beta_-^2\right)+\beta_-^2 \beta_+^2 r^2\right)\right) dt \, d\theta .
\end{align}
Now, to obtain the more familiar form of the BTZ metric \eqref{eq:BTZmetric}, we keep the time coordinate as it is and shift the angular coordinate. This also fixes the values of constants $\cC_\pm$ appearing in the Hamiltonian \eqref{eq:HamCFT} and is given by,
\begin{equation}\label{eq:timerescale}
    t\rightarrow t,\quad \theta\rightarrow \theta +\left(\frac{2  A_\phi}{r^2}\right)t,\quad \cC_\pm=\pm\frac{2}{l \, (v \pm \pi \sigma)}=\pm \frac{2}{l\,p_\pm}=\frac{\beta_\pm}{\pi \,l^2}.
\end{equation}

\subsection{Relativistic Hamiltonian}
The chemical potential $\xi_\pm$ corresponding to the relativistic hamiltonian \eqref{eq:relHam} with $W(\theta)=0$ is given by,
\begin{equation}
    \xi_\pm= \mp \cC_\pm\, p_\pm.
\end{equation}
Substituting the same form of $p_\pm$ in terms of $\beta_\pm$ as we did for the non-relativistic case, we obtain the metric for BTZ black hole,
\begin{align}
   ds^2& =\frac{\pi ^2 \, l^4 }{\beta_-^2 \beta_+^2}\left(\beta_-^2 \cC_+^2-2 \beta_- \beta_+ \cC_+ \cC_- \cosh \left(\frac{2 \rho }{l}\right)+\beta_+^2 \cC_-^2\right) dt^2  + d\rho ^2\nn\\
   &-\frac{2\pi ^2 \, l^4 }{\beta_-^2 \beta_+^2}\left(\beta_-^2 \cC_+-\beta_- \beta_+ (\cC_++\cC_-) \cosh \left(\frac{2 \rho }{l}\right)+\beta_+^2 \cC_-\right) dt\,d\theta \nn\\
   &+\frac{\pi ^2 \, l^4 }{\beta_-^2 \beta_+^2}\left(\beta_-^2+\beta_+^2-2 \beta_- \beta_+ \cosh \left(\frac{2 \rho }{l}\right)\right)d\theta^2 .
\end{align}
Similar to the non-relativistic case, we perform a change of coordinate in the $\rho$ coordinate,
\begin{equation}
     \rho=   \rho=\frac{l}{2} \cosh ^{-1}\left(\frac{\pi ^2\,l^4 (\beta_-^2 + \beta_+^2 )-\beta_-^2 \beta_+^2 r^2}{2 \pi ^2 \beta_- \beta_+ l^4}\right).
\end{equation}
This gives us metric of the form,
\begin{align}
    ds^2 &=\frac{ 1}{\beta_-^2 \beta_+^2} \left(\pi ^2 \beta_-^2 \cC_+^2 l^4+\cC_+ \cC_- \left(\beta_-^2 \beta_+^2 r^2-\pi ^2 l^4 \left(\beta_-^2+\beta_+^2\right)\right)+\pi ^2 \beta_+^2 \cC_-^2 l^4\right)dt^2\nn\\
    &+ r^2 d\theta ^2+\frac{ l^2 r^2\,\beta_-^4 \beta_+^4 }{\pi ^4 l^8 \left(\beta_-^2-\beta_+^2\right)^2-2 \pi ^2 \beta_-^2 \beta_+^2 l^4 r^2 \left(\beta_-^2+\beta_+^2\right)+\beta_-^4 \beta_+^4 r^4} dr^2 \nn\\
    &-\frac{1}{ \beta_-^2 \beta_+^2} \left(\pi ^2 \cC_+ l^4 \left(\beta_-^2-\beta_+^2\right)+\beta_-^2 \beta_+^2 \cC_+ r^2+\pi ^2 \cC_- l^4 \left(\beta_+^2-\beta_-^2\right)+\beta_-^2 \beta_+^2 \cC_- r^2\right)dt d\theta.
\end{align}
In this case, to bring to the standard BTZ metric form \eqref{eq:BTZmetric}, we perform the same $t$ and $\theta$ transformations as in \eqref{eq:timerescale} with $\cC_\pm$ chosen to be,
\begin{align}
    \cC_\pm =\pm\frac{1}{l}.
\end{align}

\section{Infinite product computation}\label{app:Infinite product computation}
In this section, we compute the infinite product, 
\begin{equation}
    \prod_{k=1}^\infty \frac{1}{1-e^{-\mu k}}
\end{equation}
Define, 
\begin{equation}
    q=e^{-\mu},\quad \text{and}\quad\tau =\frac{i\mu}{2 \pi}
\end{equation}
Hence we have, $q=e^{2\pi \,i \tau}$
Now, Dedekind eta function is defined as,
\begin{equation}\label{eq:Dedekindetafn}
    \eta(\tau)=q^{\frac{1}{24}}\prod_{k=1}^\infty(1-q^k)
\end{equation}
Hence, the product that we are concerned with can be given in terms of Dedekind eta function as,
\begin{equation}
     \prod_{k=1}^\infty \frac{1}{1-q^k}=e^{-\mu/24} \, \eta\bigg(\frac{i\mu}{2\pi}\bigg)^{-1}
\end{equation}
Since we are interested in the $\mu\rightarrow 0$ limit, it would be beneficial to use the modular transformation of Dedekind eta function,
\begin{equation}\label{eq:modulartrans}
    \eta\bigg(-\frac{1}{\tau}\bigg)=\sqrt{-i\tau}\, \eta(\tau)
\end{equation}
This implies,
\begin{equation}
    \eta\bigg(\frac{i\mu}{2\pi}\bigg) = \bigg(\frac{\mu}{2\pi}\bigg)^{-1/2}\, \eta\bigg(\frac{2\pi i}{\mu}\bigg).
\end{equation}
Now we compute $\eta\bigg(\frac{2\pi i}{\mu}\bigg)$,
\begin{align}
    \eta\bigg(\frac{2\pi i}{\mu}\bigg) &= q'^{1/24} \prod_{k=1}^\infty (1-q'^k) \nn\\
    &= e^\frac{{2\pi i(2\pi i/\mu)}}{24}=e^{\pi^2/6\mu}, 
\end{align}
because the product is equal to one, in the $\mu\rightarrow 0$ limit. Hence, the entire product becomes, 
\begin{equation}
    \prod_{k=1}^{\infty}  \frac{1}{(1-e^{-\mu k })} = \sqrt{\frac{\mu}{2 \pi}} e^{-\frac{t}{24}}e^{\frac{\pi^2}{6\mu}}.
\end{equation}
\bibliographystyle{hieeetr}
\bibliography{ads}{}

\end{document}